\documentclass[12pt]{article}

\usepackage{amssymb}
\usepackage{amsmath}
\usepackage{amscd}
\usepackage{latexsym}
\usepackage{graphicx}

\usepackage{enumitem}

\usepackage{cite}
\numberwithin{equation}{section}

\newcommand{\be}{\begin{equation}}
\newcommand{\ee}{\end{equation}}
\newcommand{\Dlt}{\Delta}
\newcommand{\dlt}{\delta}
\newcommand{\prt}{\partial}

\newcommand{\bt}{\beta}
\newcommand{\vp}{\varphi}
\newcommand{\ep}{\varepsilon}
\newcommand{\al}{\alpha}
\newcommand{\ra}{\rightarrow}
\newcommand{\sgm}{\sigma}

\newcommand{\gm}{\gamma}
\newcommand{\om}{\omega}

\newcommand{\lbd}{\lambda}

\begin{document}

\begin{center}

{\Large{\bf Selected Topics of Social Physics: \\
Nonequilibrium Systems} \\ [5mm]

Vyacheslav I. Yukalov}  \\ [3mm]

{\it
$^1$Bogolubov Laboratory of Theoretical Physics, \\
Joint Institute for Nuclear Research, Dubna 141980, Russia \\ [2mm]

$^2$Instituto de Fisica de S\~ao Carlos, Universidade de S\~ao Paulo, \\
CP 369, S\~ao Carlos 13560-970, S\~ao Paulo, Brazil} \\ [3mm]

{\bf E-mail}: {\it yukalov@theor.jinr.ru}

\end{center}

\vskip 1cm

\begin{abstract}

This review article is the second part of the project ``Selected Topics of Social 
Physics". The first part has been devoted to equilibrium systems. The present part 
considers nonequilibrium systems. The style of the paper combines the features 
of a tutorial and a review, which, from one side, makes it easy to read for 
nonspecialists aiming at grasping the basics of social physics, and from the other 
side, describes several rather recent original models containing new ideas that 
could be of interest to experienced researchers in the field. The present material 
is based on the lectures that the author had been giving during several years at 
the Swiss Federal Institute of Technology in Z\"{u}rich (ETH Z\"urich).        

\end{abstract}

\vskip 3mm
 

\newpage

{\large{\bf Contents}}

\vskip 5mm
{\bf 1. Introduction}

\vskip 3mm
{\bf 2. Dynamical Social Systems}

\vskip 2mm
   2.1. Dynamical Systems

   2.2. Stability of Solutions

   2.3. Method of Linearization

   2.4. Plane Motion

   2.5. Evolution Equations

\vskip 1mm
\hspace{6mm} 2.5.1. Differential Rate Equations

\hspace{6mm}  2.5.2. Delay Differential Equations

\vskip 1mm
   2.6. Examples of Evolution Equations

   2.7. Replicator Equation

   2.8. Free Replicator Equation

   2.9. Influence of Noise

   2.10. Fokker-Planck Equation

\vskip 3mm
{\bf 3. Generalized Evolution Equations}

\vskip 2mm
   3.1. Functional Carrying Capacity

   3.2. Evolutionary Stable States

   3.3. Punctuated Evolution

   3.4. Symbiosis of Species

\vskip 1mm
\hspace{6mm} 3.4.1. Interaction through Carrying Capacities

\hspace{6mm} 3.4.2. Uncorrelated Symbiosis

\hspace{6mm}  3.4.3. Correlated Symbiosis

\hspace{6mm}  3.4.4. Mixed Symbiosis

\vskip 1mm
   3.5. Role of Growth Rates

   3.6. Self-Organized Society

\vskip 1mm
\hspace{6mm} 3.6.1. Trait Groups

\hspace{6mm} 3.6.2. Coexistence of Cooperators and Defectors

\hspace{6mm} 3.6.3. Three Coexisting Groups 

\vskip 3mm
{\bf 4. Models of Financial Markets}

\vskip 2mm
   4.1. Efficient Market Model

   4.2. Diffusion Price Model

   4.3. Herding Market Model

   4.4. Time Series Analysis

\vskip 3mm
{\bf 5. Conclusion}

\vskip 3mm

\section{Introduction}

The term social physics was introduced by Comte, who gave the definition 
\cite{Iggers_1959}: {\it Social physics is that science which occupies itself 
with social phenomena, considered in the same light as astronomical, physical, 
chemical, and physiological phenomena, that is to say as being subject to natural 
and invariable laws, the discovery of which is the special object of its researches}. 
Comte \cite{Comte_1975} divided sociology into two main fields, or branches: social 
statics, or the study of the forces that hold society together; and social dynamics, 
or the study of the causes of social change. For social dynamics, Comte also used 
the term {\it social evolutionism}. The term "social physics" was widely used by 
Quetelet \cite{Quetelet_1835}.

The whole review consists of two parts. The first part \cite{Yukalov_0} is devoted 
to static, or equilibrium, phenomena and systems. In the present part of the review, 
dynamic, or nonequilibrium, phenomena and systems are considered. This implies that 
the considered social systems have to be characterized by evolution equations describing 
how the systems develop and what are their evolutionary stable states.  

It goes without saying that in a rather short review article it is impossible to cover
the whole of social physics, being a highly multidisciplinary science including a 
variety of many different topics. Applications of mathematical and physical methods 
to social systems are really extremely ramified. As examples, it is possible to 
mention the growth and distribution of cities 
\cite{Zipf_1949,Makse_1995,Soo_2005,Newman_2005,Batty_2008,Saichev_2010,Batty_2013,
Barthelemy_2016,Barthelemy_2018,Barthelemy_2019,Bettencourt_2021}, 
dynamics of urban population \cite{Stouffer_1940,Gonzalez_2008,Park_2010,Lee_2015}, 
optimization of city traffic 
\cite{Leutzbach_1988,Kessels_1999,Chowdhury_2000,Helbing_2001,Nagatani_2002,Ni_2015,
Kerner_2017}, econophysics 
\cite{Mantegna_1999,Sornette_2003,Bouchaud_2003,Watanabe_2009,Aoyama_2010,Takayasu_2010,
Andersen_2013,Takayasu_2013,Ormerod_2016,Abergel_2017,Slanina_2018,Bouchaud_2018},  
the evolution of cooperation, and, more generally, interplay between cooperation and 
competition 
\cite{Dawes_1980,Smith_1982,Weibull_1997,Roca_2009,Perc_2010,Tanimoto_2015,Wang_2015,
WangWang_2015,Su_2018,Henrich_2021},   
separation of societies into subgroups and communities 
\cite{Gronlund_2004,Masuda_2010,Perc_2012,Chen_2014},  
voter models \cite{Suchecki_2005,Mobilia_2007,Galam_2008,Masuda_2010,Redner_2019},
human-like artificial intelligence 
\cite{Russell_2002,Bengio_2013,Goodfellow_2016,Floridi_2019,Liang_2019,Theodorou_2020,
Bachrach_2020,Bonnefon_2020,Yukalov_2023},
criminology \cite{Becker_1968,Cohen_1979,Bardhan_1997,Johnson_2004,Zimring_2006,
Nuno_2011,Ball_2012,Dorsogna_2015,Szolnoki_2017}, 
migration \cite{Sullivan_1993,Borjas_1994,Kunovich_2013}, 
contagion phenomena and the spread of infections 
\cite{Kermack_1927,Hufnagel_2004,Salathe_2018},
ecological problems \cite{Lindeman_1942,Scheffer_2000,Roos_2001}, climate change 
\cite{Richardson_1922,Tsonis_2004,Gladwell_2006,Ranson_2014}, various complex networks
\cite{Wasserman_1994,Albert_2002,Dorogovtsev_2003,Watts_2003,Pastor_2004,Barrat_2008,
Lewis_2009,Newman_2010,Estrada_2011,Caldarelli_2012,Sen_2014,Latora_2017,Manoj_2018},
and many other topics \cite{Perc_30,Perc_31,Jusup_32}.    

Since it is impossible to embrace the immensity, it is necessary to select some of 
topics. The choice of topics in the present review is dictated by the main aims
that are: First, to give an introduction into the methods of dealing with dynamical 
systems that are necessary for analysing the behavior of social systems. Second, to
illustrate these methods on simple known models. The last, but not the least, in order
to make the present review original, it was necessary to consider some novel, fresh
models that have not been mentioned in other review articles. It was natural to choose 
the examples of new models from those fields that are in the frame of the author's 
research interests. Thus the main difference of the present article from the previously 
published surveys is in two points: (i) More detailed exposition of the methods used 
for treating social systems. (ii) The consideration of new models that have not been 
studied in the previous reviews. 
 
The exposition starts with a brief survey of dynamical systems and an overview of 
the main known evolution equations. Then several recent models are considered 
allowing for the description of nontrivial dynamical effects. Evolution equations 
with functional carrying capacity are introduced. Punctuated evolution is described. 
A new type of evolution equations characterizing symbiosis is presented. The 
peculiarity of constructing the evolution equations for structural self-organized 
societies is discussed. Some examples of financial markets are treated, emphasizing 
the important role of herding behavior.

\section{Dynamical Social Systems}

The evolution of nonequilibrium social systems is described by the time dependence 
of observable quantities. The time dependence is usually given through differential 
equations. Sometimes, one considers discrete time, when the evolution equations 
are given by difference equations. Generally, time is a continuous variable. Here we 
shall mainly deal with differential equations.

\subsection{Dynamical Systems}

The evolution of social systems is described by considering the temporal evolution 
of observable quantities. As examples of observable quantities, it is possible to 
mention order parameters, the value of total order, and the fractions of populations. 
The system of differential equations characterizing the system evolution, from the 
mathematical point of view, is a dynamical system which many books are devoted to 
(see, e.g., \cite{Nemytsky_41,Rudolph_42,Robinson_43,Mikhailov_44,Orlando_45}). The 
main definitions we shall need are briefly surveyed below. 

Let the set of observable quantities of interest be $x_1, x_2,\ldots, x_N$. Each of 
the observables varies inside the related manifold $\mathbb{X}_n$, which is a part 
of the real-number axis $\mathbb{R} = (-\infty, +\infty)$. In order not to overload 
the text by notations, we shall simply write that $x \in \mathbb{R}^N$. 

The set
\be
\label{4.1}
 x \equiv \{ x_n : ~ n = 1,2, \ldots, N\} \in \mathbb{R}^N
\ee
is the {\it system state}. The manifold $\mathbb{R}^N$ can be termed {\it phase space}.
The time arrow is the variation of time $t$, beginning at an initial moment $t_0$ and 
increasing to $\infty$. Without the loss of generality, it is usually admissible to 
set $t_0=0$. So, often time is treated as the variable in the interval $[0,\infty)$.

The {\it evolution law} or the {\it law of motion} is the dependence on time $t$ of 
the variables $x_n$. The time dependent variable
\be
\label{4.2}
 x = x(t,x_0) \;  ,
\ee
with a state $x_0$ given at a moment of time $t_0$, called {\it initial state},
\be
\label{4.3}
 x_0 = x(t_0,x_0) = \{ x_{0n} : ~ n = 1,2,\ldots,N\} \; ,
\ee
is a {\it dynamical state}.

The family of evolution equations forms a {\it general dynamical system}. Often, 
one also requires that the states of a dynamical system would satisfy the group 
composition law
\be
\label{4.4}
x(t+t',x_0) = x(t,x(t',x_0) ) \;  .
\ee
The evolution law (\ref{4.2}), with the group property (\ref{4.4}), consists of the 
endomorphisms projecting the phase space onto itself. The family of the endomorphisms
composes a {\it flow} 
\be
\label{4.5}
\{ x(t,\ldots) : ~ 
\mathbb{R}_+ \times \mathbb{R}^N \longmapsto \mathbb{R}^N \} \; .
\ee
The set of all states forms a {\it trajectory}
\be
\label{4.6}
\{ x(t,x_0): ~ t \geq t_0 \} \; .
\ee
The states at fixed times are called the points of the trajectory.
    
General dynamical systems in continuous time are usually represented by differential 
equations in the normal form
\be
\label{4.7} 
 \frac{dx}{dt} = f(x,t) \; ,
\ee
where
\be
\label{4.8}
 f(x,t) = \{ f_n(x,t) : ~ n = 1,2,\ldots, N\} \; .
\ee
The differential equations containing higher-order time derivatives can always be 
reduced to the normal form by means of renotations. The dynamical system dimension 
is the number of the evolution equations. In the present case, it is $N$.

If the right-hand side $f(x,t)$ explicitly depends on time, then (\ref{4.7}) is 
termed a nonautonomous equation, while when $f(x,t) = f(x)$ does not depend on time, 
this is an autonomous equation. The intermediate case is when at finite times the 
evolution equation is nonautonomous, but, as time increases, the equation becomes 
autonomous, so that
$$
 \lim_{t\ra\infty} f(x,t) = f(x) \; .
$$
Then the equation is called transiently nonautonomous. This case corresponds to a 
nonequilibrium system subject to the action of external forces perturbing the system 
during a finite period of time. 

A {\it fixed point} or stationary point is a state
\be
\label{4.9}
 x^* = \{ x_n^*: ~ n = 1,2, \ldots, N\} \; ,
\ee
for which there exists a time $t^*$, such that 
\be
\label{4.10}
 x(t,x^*) = x^* \qquad ( t \geq t^*) \; .
\ee

The norm of a dynamic state is denoted as
\be
\label{4.11}
|| x(t,x_0) || = \sqrt{ \sum_{n=1}^N x_n^2(t,x_0) } \; ,
\ee
which is termed {\it vector norm} or Euclidean norm. This norm describes the 
character of motion that can be singular, or collapsing, and it can be nonsingular, 
or noncollapsing.

The motion, starting at a point $x_0$, is strongly singular, if there exists such 
a finite moment of time $t_c=t_c(x_0)$, called critical, when the trajectory diverges 
to infinity, 
\be
\label{4.12}
 || x(t,x_0) || \ra \infty \qquad ( t \ra t_c) \;  .
\ee
The motion is weakly singular, when the trajectory diverges as time goes to infinity:
\be
\label{4.13}
 || x(t,x_0) || \ra \infty \qquad ( t \ra \infty) \; .
\ee
The motion is nonsingular, if the trajectory is bounded and never diverges.

\subsection{Stability of Solutions}

Solutions to evolution equations can be stable or unstable \cite{Lyapunov_46}. Finding 
the regimes of stability for the solutions to the equations of motion is one of the 
main problems in studying the evolution of social systems. There are several types of 
stability. The most general type of stability is the Lagrange stability.

\vskip 2mm
{\it Lagrange Stability}. The motion, starting at a point $x_0$, is Lagrange stable, 
if it is nonsingular for all times, that is, the trajectory is always bounded,
\be
\label{4.14}
 || x(t,x_0) || \; < \; \infty \qquad ( t \geq t_0 ) \;  .
\ee

\vskip 2mm
{\it Global Lagrange Stability}. The motion is globally Lagrange stable, if it is 
Lagrange stable for all initial conditions, so that
\be
\label{4.15}
 || x(t,x_0) || \; < \; \infty \qquad ( x_0 \in \mathbb{R}^N\; , ~ t \geq t_0 ) \; .
\ee

\vskip 2mm 
{\it Poincar\'{e} Stability}. The solution $x(t,x_0)$ is Poincar\'e stable if, for 
any $\ep>0$, there exist $\dlt_\ep$ and a time $t_\varepsilon$, called recurrence time, 
such that, if $| x_0 - x_0'|< \dlt_\ep$, then 
\be
\label{4.16}
 || x(t_\ep,x_0') - x_0 || \; < \; \ep \;   .
\ee
Since any moment of time can be accepted as a new initial point, there can exist an 
infinite sequence or recurrence times. 

\vskip 2mm

{\it Lyapunov Stability}. The solution $x(t,x_0)$ is Lyapunov stable if, for any 
$\ep>0$, there exists $\dlt_\ep$ such that, if $\Vert x_0 - x_0'\Vert < \dlt_\ep$, 
then for all $t>0$
\be
\label{4.17}
 || x(t,x_0) - x(t,x_0') || \; < \; \ep \qquad ( t > 0 ) \; .
\ee
This means that if two trajectories, at the initial time, start sufficiently close 
to each other, they remain close to each other for all times $t>0$. 

\vskip 2mm

{\it Global Lyapunov Stability}. The solution $x(t,x_0)$ is globally Lyapunov stable, 
if it is Lyapunov stable for any initial condition $x_0\in\mathbb{R}^N$.

\vskip 2mm
{\it Asymptotic Lyapunov Stability}. The solution $x(t,x_0)$ is asymptotically stable, 
if it is Lyapunov stable and there exists $\delta_\ep > 0$ such that if
$\Vert x_0 - x_0'\Vert < \dlt_\ep$, then
\be
\label{4.18}
\lim_{t\ra\infty} || x(t,x_0) - x(t,x_0') || = 0 \;  .
\ee
That is, if two asymptotically stable trajectories, at the initial time, start 
sufficiently close to each other, they coincide for the time tending to infinity.

\vskip 2mm
{\it Global Asymptotic Stability}. The solution $x(t,x_0)$ is globally asymptotically 
stable, if it is asymptotically stable for any initial condition $x_0\in\mathbb{R}^N$.

\vskip 2mm
Definitions of stability can be straightforwardly reformulated for fixed points. For 
the autonomous equations, fixed points $x^*$ are given by the condition
\be
\label{4.19}
 \frac{dx^*}{dt} = f(x^*) = 0 \;  .
\ee

\vskip 2mm
{\it Lyapunov Stable Fixed Point}. The fixed point $x^*$ is Lyapunov stable, if for 
any $\ep>0$ there exists $\dlt_\ep>0$ such that, if $\Vert x_0-x^*\Vert < \dlt_\ep$, 
then 
\be
\label{4.20}
 || x(t,x_0) - x^* ||  < \ep \qquad ( t \geq 0 )
\ee
for all $t>0$. 

\vskip 2mm
{\it Globally Lyapunov Stable Fixed Point}. The fixed point $x^*$ is globally Lyapunov 
stable if it is Lyapunov stable for all initial conditions $x_0 \in \mathbb{R}^N$.

\vskip 2mm
{\it Asymptotically Stable Fixed Point}. The fixed point $x^*$ is asymptotically 
stable, if it is Lyapunov stable and there exists $\dlt>0$ such that, if 
$\Vert x_0 - x^* \Vert < \dlt$, then
\be
\label{4.21}
 \lim_{t\ra\infty} || x(t,x_0) - x^* || = 0 \;  .
\ee

\vskip 2mm
{\it Globally Asymptotically Stable Fixed Point}. The fixed point $x^*$ is globally 
asymptotically stable if it is asymptotically stable for all initial conditions 
$x_0 \in \mathbb{R}^N$.

\vskip 2mm
An asymptotically stable fixed point is called an {\it attractor}, since it 
attracts all trajectories starting in the vicinity of an initial point $x_0$. 
The maximal region in the phase space, around an initial point $x_0$, from where 
all trajectories tend to an attractor $x^*$, is termed the {\it basin of attraction}. 
The motion is called regular, when it is Lyapunov stable, and the motion is 
chaotic, when it is Lyapunov unstable.

\subsection{Method of Linearization}

Lyapunov developed two methods for analyzing the stability of motion. The most often 
used is the linearization method that we briefly remind below. 

The analysis of stability starts with considering a small deviation $\delta x(t)$ 
from the evolution of the variable $x(t)$:
\be
\label{4.22}
x(t) \longmapsto x(t) + \dlt x(t) \; ,
\ee
where the deviation is a vector
\be
\label{4.23}
\dlt x(t) = \{ \dlt x_n(t) : ~ n = 1,2,\ldots, N\} \;  .
\ee
Substituting this into the evolution equation, we keep in mind that $x(t)$ is a 
solution to this equation. Taking into account only the linear deviation, we get 
\be
\label{4.24}
 \frac{d}{dr} \; \dlt x(t) = \hat J(x,t) \dlt x(t) \; ,
\ee
with the Jacobian matrix
\be
\label{4.25}
 \hat J(x,t) = [\; J_{mn}(x,t) \; ] \; , \qquad
J_{mn} \equiv \frac{\prt f_m(x,t)}{\prt x_n} \;  .
\ee

We can define the eigenvalues and eigenvectors of the Jacobian matrix by the 
eigenproblem
\be
\label{4.26}
 \hat J(x,t) \vp_n(x,t) = J_n(x,t) \vp_n(x,t) \;  ,
\ee
in which the eigenvectors are the vector-columns
\be
\label{4.27}
 \vp_n(x,t) = \{ \vp_{nm}(x,t): ~ m = 1,2, \ldots, N\} \; .
\ee
Explicitly, equation (\ref{4.26}) can be written in the form
$$
 \sum_m (J_{nm} - J_n \dlt_{nm}) \vp_{nm} = 0 \; ,
$$
where, for short, the variables are not shown. This equation possesses a nontrivial
solution, provided that the eigenvalues are given by the condition of the zero 
determinant
\be
\label{4.28}
 | \; \hat J(x,t) - J_n(x,t) \; | = 0 \;  .
\ee

The linearization is usually done in the vicinity of fixed points, provided these
exist, for which it is required that the evolution equation in the limit of large 
time becomes autonomous,
\be
\label{4.29}
 f(x,t) \simeq f(x) \qquad ( t \ra \infty )  \; .
\ee
Then, the Jacobian, as well as its eigenvalues and eigenvectors, in the limit of 
large time, do not depend on time:
\be
\label{4.30}
\hat J(x,t) \simeq \hat J(x) \; , \qquad \vp_n(x,t) \simeq \vp_n(x) 
\qquad
( t \ra \infty )  \;  .
\ee

Considering small deviations
\be
\label{4.31}
\dlt x(t) = x(t) - x^*
\ee
from a fixed point, given by the definition
\be
\label{4.32}
  f(x^*) = 0 \; ,
\ee
we have the equation  
\be
\label{4.33}
 \frac{d}{dt} \; \dlt x(t) = \hat J(x^*) \; \dlt x(t) \;  .
\ee
The Jacobian eigenproblem takes the form
\be
\label{4.34}
 \hat J(x^*) \vp_n(x^*) = J_n(x^*) \vp_n(x^*) \;  .
\ee
The solution to the linear equation for the deviation can be represented as 
an expansion over the eigenvectors of the Jacobian matrix:
\be
\label{4.35}
 \dlt x(t) = \sum_n c_n \exp\{ J_n(x^*) t \} \vp_n(x^*) \; .
\ee

The {\it Lyapunov exponents}, or characteristic exponents, related to a fixed 
point, are the real parts of the Jacobian eigenvalues evaluated at this fixed 
point,
\be
\label{4.36}
 \lbd_n \equiv {\rm Re} J_n(x^*) \;  .
\ee
The collection $\{\lbd_n\}$ of all Lyapunov exponents forms the {\it Lyapunov 
spectrum}. The number of Lyapunov exponents equals the dimension of the phase 
space. The largest Lyapunov exponent from the Lyapunov spectrum is called 
the {\it convergence rate},
\be
\label{4.37}
 \lbd \equiv \sup_n \lbd_n \;  .
\ee
Lyapunov exponents characterize the asymptotic stability of fixed points and 
describe the type of the asymptotic dynamics. The sign of the largest Lyapunov 
exponent (\ref{4.37}) determines the asymptotic convergence or divergence of 
trajectories with infinitesimally close initial conditions. If the largest 
exponent is negative, the motion is regular, so that the trajectories with 
infinitesimally close initial conditions converge to each other. But if the 
largest exponent is positive, then the initially close trajectories diverge, 
and the dynamical system is chaotic. More information on chaotic motion can be 
found in literature \cite{Ruelle_1989,Lorentz_1993,Zaslavskii_2009}.

\vskip 2mm

{\bf Lyapunov theorem}. {\it If all Lyapunov exponents for a fixed point are
negative, this fixed point is asymptotically stable. And if at least one of the 
Lyapunov exponents is positive, this fixed point is asymptotically unstable}.

\vskip 2mm
As a simplest case, let us consider a one-dimensional autonomous dynamical system. 
The Jacobian is just the derivative
$$
J(x) = \frac{\prt f(x)}{\prt x} \;   .
$$
For each fixed point, there is a single Lyapunov exponent 
$$
 \lbd = {\rm Re} J(x^*) \;  .
$$
The fixed point is asymptotically stable, if $\lambda < 0$. Then, there exists a 
basin of attraction from which the trajectories tend to the given fixed point.
The fixed point is termed neutrally stable, when $\lambda = 0$. In this situation, 
the fixed point can be a center, around which the trajectory oscillates. The fixed 
point is asymptotically unstable, when $\lambda > 0$. Then, there is no basin of 
attraction for this fixed point, but all trajectories diverge from it.

\subsection{Plane Motion}

The motion on a plane is described by a two-dimensional dynamical system. That is, 
the phase space is $\mathbb{R}^2$. For an autonomous system, there are two 
differential equations
\be
\label{4.38}
 \frac{dx}{dt} = f_1(x,y) \; , \qquad  \frac{dy}{dt} = f_2(x,y) \; ,
\ee
with initial conditions
\be
\label{4.39}
 x(0) = x_0 \; , \qquad y(0) = y_0 \;  .
\ee

Dividing the second equation over the first gives the relation
\be
\label{4.40}
\frac{dy}{dx} = \frac{f_2(x,y)}{f_1(x,y)}
\ee
defining the trajectory
\be
\label{4.41}
 y = y(x,x_0,y_0) \;  .
\ee
The set of the trajectories in the plane for different initial conditions forms 
the {\it phase portrait} of the dynamical system
\be
\label{4.42}
 \{ y(x,x_0,y_0): ~ x_0 \in \mathbb{R}^2 \; , ~ y_0 \in \mathbb{R}^2 \} \; .
\ee

The Jacobian is a two-by-two matrix with the elements
$$
J_{11} = \frac{\prt f_1}{\prt x} \; , \qquad 
J_{12} = \frac{\prt f_1}{\prt y} \; ,
$$
\be
\label{4.43}
J_{21} = \frac{\prt f_2}{\prt x} \; , \qquad 
J_{22} = \frac{\prt f_2}{\prt y} \;   ,
\ee
where $J_{mn} \equiv J_{mn}(x,y)$, but, for brevity, the dependence on $x$ and $y$ 
is omitted. The Jacobian eigenvalues are given by the equation
\begin{eqnarray}
\label{4.44}
\left| \begin{array} {ll}
J_{11} - J ~ & ~ J_{12} \\
J_{21} ~ & ~ J_{22} - J \end{array} \right| 
= 0 
\end{eqnarray}  
resulting in two solutions
\be
\label{4.45}
 J_{1,2} = \frac{1}{2} \; \left[ \; {\rm Tr} \hat J \pm 
\sqrt{ ({\rm Tr} \hat J )^2 - 4 {\rm det} \hat J } \; \right] \; ,
\ee
in which
$$
 {\rm Tr} \hat J = J_{11} + J_{22} \; , \qquad
{\rm det} \hat J = J_{11} J_{22} - J_{12} J_{21} \;  .
$$

The solutions for the fixed points, given by the equations
\be
\label{4.46}
 f_1(x^*,y^*) = 0 \; , \qquad  f_2(x^*,y^*) = 0 \;  ,
\ee
can be of the following types depending on the Jacobian eigenvalues 
$J_n = J_n(x^*,y^*)$.  

\vskip 2mm
{\it Stable node}: $J_1 \leq J_2 < 0$, hence $\lambda_1 \leq \lambda_2 < 0$. 
There exists a basin of attraction, from which all trajectories tend to this 
fixed point without oscillations.

\vskip 2mm
{\it Unstable node}: $J_1 \geq J_2 > 0$, hence $\lambda_1 \geq \lambda_2 > 0$. 
There is no basin of attraction, and all trajectories, starting in the vicinity 
of this fixed point diverge without oscillations. 

\vskip 2mm
{\it Stable focus}: $J_1 = \lambda + i \omega$ and $J_2 = \lambda - i \omega$,
with $\lambda < 0$ and $\omega > 0$. There is a basin of attraction, from which 
all trajectories tend to the fixed point, oscillating around it.

\vskip 2mm
{\it Unstable focus}: $J_1 = \lambda + i \omega$ and $J_2 = \lambda - i \omega$,
with $\lambda > 0$ and $\omega > 0$. There is no basin of attraction, and all 
trajectories, starting in the vicinity of the fixed point, diverge from it, 
oscillating around it.

\vskip 2mm
{\it Elliptic point}: $J_1 = i \omega$ and $J_2 = -i \omega$, with $\omega > 0$.
There exists a limit cycle, either stable or unstable. A stable limit cycle is 
a closed orbit, surrounding the fixed point, and attracting all trajectories 
starting from the related basin of attraction. An unstable limit cycle is a closed 
orbit, surrounding the fixed point, and repelling all trajectories, starting in its
vicinity.

\vskip 2mm
{\it Saddle point}: $J_1 < 0$ and $J_2 > 0$, hence $\lambda_1 < 0 < \lambda_2$.
There are only two trajectories tending to the point, while all other trajectories 
diverge from it.

\vskip 2mm

For continuous dynamical systems on the plane, there exists the Poincare-Bendixson 
theorem showing that such systems enjoy rather regular motion and do not exhibit
chaotic behavior.  

\vskip 2mm

{\bf Poincare-Bendixson theorem}. {\it If a trajectory of a continuous 
two-dimensional dynamical system is bounded, that is, Lagrange stable, then it 
approaches either a fixed point or a limit cycle}.

\subsection{Evolution Equations}

Evolution equations are usually represented by differential equations and sometimes
by delay differential equations. Below we mention the general structure of such 
equations and then specify these equations by several examples.

\subsubsection{Differential Rate Equations}

Social systems are composed of population groups whose members interact with each 
other cooperating or competing. The number of members in each group can vary, which
can be characterized by differential equations.  
 
Let us consider a society composed of several different groups of populations
enumerated by the index $i = 1,2,\ldots$. In an $i$-th group there are $N_i$
members. The number of the members $N_i = N_i(t)$ is not fixed but can vary, 
e.g. because of the members births and deaths, as well as due to newcomers 
joining the society from outside. The total number of the society members is
\be
\label{4.47}
 N = \sum_i N_i = N(t) \;  .
\ee
In other cases, $N$ can represent a company capitalization, with $N_i$ being the
capitalization of the company parts.

The number of the society members can be very large because of which it is more
convenient to use reduced numbers, for example normalizing $N_i$ by a fixed scaling 
$N_0$, so that 
\be
\label{4.48}
 x_i \equiv \frac{N_i}{N_0} \qquad ( i = 1,2, \ldots ) \;  .
\ee
Often $N_0$ can be chosen as the total number of the society members at the initial 
time, $N(0)$, however this is not compulsory. The scaling number is chosen according
to the convenience and can be different for different cases. The set 
$x_i:\; i=1,2,\ldots$ of all variables forms the society state.
 
The evolution equations for the society groups usually have the form of the rate 
equation
\be
\label{4.49}
 \frac{dN_i}{dt} = R_i N_i + \Phi_i \; ,
\ee
where $R_i = R_i(x,t)$ is the effective rate of the population group change and 
$\Phi_i = \Phi_i(x,t)$ is an external population flux. When the rate $R_i$ is positive,
the group population increases, while if the rate $R_i$ is negative, the group
population decreases.  

The effective rate is often taken in the form
\be
\label{4.50}
 R_i = \gm_i + \sum_j A_{ij} N_j \;  ,
\ee
where $\gamma_i$, e.g. is a birth (death) rate and the second term reflects the 
influence on the rate of other groups. The equation for the number of the group 
members reads as
\be
\label{4.51}
 \frac{dN_i}{dt} = \left( 
\gm_i + \sum_j A_{ij} N_j \right) N_i + \Phi_i \;  ,
\ee
and in the reduced form the evolution equation becomes
\be
\label{4.52}
\frac{dx_i}{dt} = \left( 
\gm_i + \sum_j a_{ij} x_j \right) x_i + \vp_i \; ,
\ee
with the notation
$$
a_{ij} \equiv A_{ij} N_0 \; , \qquad 
\vp_i \equiv \frac{\Phi_i}{N_0} \; .
$$

\subsubsection{Delay Differential Equations}

Sometimes, modeling the effective rates of evolution equations, one takes into 
account that the description of realistic processes involves delayed actions, 
so that the evolution equation for a species $x_i$ becomes delay-differential 
equation \cite{Gopalsamy_1992,Kuang_1993}. The delay differential equations belong 
to the class of functional equations, similarly to partial differential equations, 
which are infinite dimensional \cite{Temam_1988}.

In the reduced form, the delay equation for the vector
$$
x(t) = \{ x_n(t) : ~ n = 1,2, \ldots \} \;   ,
$$
with different discrete delays $\tau_n$ for each population, reads as
\be
\label{d.1}
 \frac{d}{dt} \; x(t) = f(x(t),x_1(t-\tau_1),x_2(t-\tau_2),\ldots, t) \; ,
\ee
where $\tau_n$ are delay times and $f=\{f_n\}$ is also a vector. The history 
conditions for negative times are to be given by prescribed history functions
\be
\label{d.2}
x_n(t) = h_n(t) \qquad ( t \leq 0 ) \;   ,
\ee
with $h_n(t)$ being explicit functions of time. 

If for each species, the delay time $\tau_n = \tau$ is the same, the evolution 
equation takes the form
\be
\label{d.3}
  \frac{d}{dt} \; x(t) = f(x(t),x(t-\tau),t) \;  .
\ee
 
The solution of a delay equation can be done following the step-by-step method. 
For instance, let us consider the case of a single population described by 
equation (\ref{d.3}), with the history condition $x(t) = h(t)$ for $t \leq 0$.
One considers the solution separately in the subsequent time intervals $[0,\tau]$, 
then $[\tau, 2 \tau]$, then $[2 \tau, 3 \tau]$, etc. At the first step, the solution
\be
\label{d.4}
x(t) = x^{(1)}(t) \qquad ( 0 \leq t \leq \tau )
\ee
is defined by the equation
\be
\label{d.5}
  \frac{d}{dt} \; x^{(1)}(t) = f(x^{(1)}(t),h(t),t) \; ,
\ee
with the initial condition
\be
\label{d.6}
x^{(1)}(0) = h(0) \; .
\ee

At the second step, denoting
\be
\label{d.7}
  x(t) = x^{(2)}(t) \qquad ( \tau \leq t \leq 2\tau) \;  ,
\ee
one finds the solution from the equation
\be
\label{d.8}
\frac{d}{dt} \; x^{(2)}(t) = f(x^{(2)}(t),x^{(1)}(t),t) 
\ee
with the initial condition
\be
\label{d.9}
x^{(2)}(\tau) = x^{(1)}(\tau) \;   .
\ee

Continuing this procedure for subsequent time intervals, at the step $k$, the solution
\be
\label{d.10}
x(t) = x^{(k)}(t) \qquad ( ( k-1)\tau \leq t \leq k\tau )
\ee
is given by the equation
\be
\label{d.11}
\frac{d}{dt} \; x^{(k)}(t) = f(x^{(k)}(t),x^{(k-1)}(t),t) \;   ,
\ee 
with the initial condition
\be
\label{d.12}
 x^{(k)}((k-1)\tau) = x^{(k-1)}( ( k-1)\tau ) \;  .
\ee
In practice, the solution at each step is calculated numerically.

\subsection{Examples of Evolution Equations}

There exist many variants of evolution equations. Below, we briefly mention several 
the most known equations. 

\vskip 2mm 

{\bf Malthus Equation}
 
\vskip 2mm

Malthus \cite{Malthus_47} was interested in the Earth population growth. He noticed 
that the Earth population varies according to the law
\be
\label{4.53}
 \frac{dN}{dt} = \gm N \;  .
\ee
The variation rate is the difference between the growth, $\gamma_+$, and death, $\gm_-$, 
rates of the population,
\be
\label{4.54}
  \gm = \gm_+ - \gm_- \; .
\ee
Taking for the scaling $N_0 = N(0)$, we get the equation in the reduced form
\be
\label{4.55}
 \frac{dx}{dt} = \gm x \;  ,
\ee
with the initial condition 
\be
\label{4.56}
x(0) = 1.
\ee
This gives the evolution law
\be
\label{4.57}
x = e^{\gm t} \;   .
\ee

As is evident, with time, the population either diminishes to zero, if $\gm< 0$, 
or tends to infinity, if $\gm>0$. Since for the people on the Earth, the birth rate 
surpasses the death rate, hence $\gm>0$, the population will grow to infinity. The 
singular behavior of the population results in overpopulation, which is called the 
Malthusian catastrophe. The population explosion leads to wars, epidemics, and hunger. 
To avoid overpopulation, Malthus advised to keep $\gamma$ equal to zero.

\vskip 2mm

{\bf Neo-Malthusian Catastrophe}

\vskip 2mm

Moreover, some authors have suggested that the situation is even worse than predicted 
by Malthus, and the Earth population can become strongly singular, diverging at a finite 
moment of time. This situation can be compared with the proliferation of cancer cells 
killing an organism at finite time \cite{Hern_48}.

The finite-time divergence can happen when, in addition to the birth rate, one 
includes into the effective rate the influence of population cooperation, so that the 
rate becomes
\be
\label{4.58}
R = \gm + AN \qquad ( \gm > 0 \; , ~ A > 0 ) \;  .
\ee
Then the evolution equation reads as
\be
\label{4.59}
  \frac{dN}{dt} = (\gm + A N) N \; .
\ee
With the scale $N_0 = N(0)$, we have the reduced equation
\be
\label{4.60}
   \frac{dx}{dt} = (\gm + a x) x \;  .
\ee

The solution
\be
\label{4.61}
x = \frac{\gm e^{\gm t}}{\gm - a(e^{\gm t}-1)}
\ee
shows that the population diverges as
\be
\label{4.62}
x \simeq \frac{1}{a(t_c-t)} \qquad ( t \ra t_c - 0 ) \;  ,
\ee
at the finite time
\be
\label{4.63}
 t_c = \frac{1}{\gm}\; \ln \left( 1 + \frac{\gm}{a} \right) \; .
\ee
For the Earth population, this time corresponds to the year 2027. Thus the human 
population is analogous to cancer cells destroying their habitat.

\vskip 2mm

{\bf Logistic Equation}

\vskip 2mm

Verhulst \cite{Verhulst_49} suggested that no population can grow indefinitely, 
but there exists a carrying capacity $K$, above which the population growth
is impossible. He proposed the logistic equation
\be 
\label{4.64}
\frac{dN}{dt} = \gm \left( 1 \; - \; \frac{N}{K} \right) N \qquad
(\gm > 0 \; , ~ K > 0 )
\ee
taking into account the existence of the carrying capacity $K$ entering the term 
characterizing the competition between the society members. The Malthus equation 
can be valid only at the very beginning of the population growth when $N \ll K$.

The logistic equation has found a number of applications describing the growth of 
population in demography, company capitalization in economics, neuron signals in 
neural networks, tumor size in medicine, reactant mass in chemistry, and so on. The 
reduced form of the equation, with the population normalized to $N_0=N(0)$, is
\be
\label{4.65}
 \frac{dx}{dt} = ( \gm  - ax) x  \; ,
\ee
where
$$
 a \equiv \gm \; \frac{N_0}{K} ~ > ~ 0 \;  .
$$
The solution to the logistic equation is the logistic function
\be
\label{4.66}
 x = \frac{\gm e^{\gm t}}{\gm + a(e^{\gm t}-1)} \;  ,
\ee
also called sigmoid function. Its limit with respect to time is finite,
\be
\label{4.67}
 \lim_{t\ra \infty}\; x = \frac{\gm}{a} = \frac{K}{N_0} \qquad ( \gm > 0 ) \;  ,
\ee
hence there is no singular behavior.    

From the point of view of the Lyapunov analysis, fixed points are defined by 
the equation
$$
 ( \gm  - ax^*) x^*  = 0
$$
resulting in two fixed points, $x_1^* = 0$ and $x_2^* = \gamma/a$. The Jacobian 
$$
J(x) = \gm - 2ax 
$$
takes the values 
$$
J(x_1^*) = \gm \; , \qquad J(x_2^*) = - \gm \; .
$$
Hence the trivial fixed point $x_1^* = 0$ is not stable, but the nontrivial fixed 
point $x_2^* > 0$ is stable.

\vskip 2mm

{\bf Lotka-Volterra Model}

\vskip 2mm

The model \cite{Lotka_50,Volterra_51} describes the coexistence of two populations, 
where one of them corresponds to prey ($x$) and the other, to predators ($y$), because 
of which it is also named the predator-prey model. This model is also employed in many 
other applications. For example, in combustion theory these can be a passive radical 
($x$) and an active radical ($y$), in medicine, it can be susceptible to infection 
individuals ($x$) and infective individuals ($y$), in economics, buying population 
($x$) and sold goods ($y$). 

In the reduced form, the corresponding equations are
\be
\label{4.68}
 \frac{dx}{dt} = \gm_1 x - a_{12} yx \; , 
\qquad
 \frac{dy}{dt} = - \gm_2 y + a_{21} xy \;  ,
\ee
where all parameters are assumed to be positive. As usual, the initial conditions
are denoted as 
\be
\label{4.69}
  x(0) = x_0 \; , \qquad y(0) = y_0 \; .
\ee
When $x_0=0$, then $y(t)$ dies down to zero as time increases. If $y_0=0$, then 
$x(t)$ explodes to infinity with time. More interesting are the nonzero initial 
conditions.
 
Let us employ the Lyapunov stability analysis. The fixed-point equations
\be
\label{4.70}
 ( \gm_1 - a_{12} y^* ) x^* = 0 \; , \qquad
  ( - \gm_2 + a_{21} x^* ) y^* = 0 \; ,
\ee
yield two fixed points: one is trivial
\be
\label{4.71}
x_1^* = 0 \; , \qquad y_1^* = 0 \; ,
\ee
and the other is nontrivial,
\be
\label{4.72}
 x_2^* = \frac{\gm_2}{a_{21}} \; , \qquad
 y_2^* = \frac{\gm_1}{a_{12}} \; .
\ee

The Jacobian matrix is composed of the elements
\be
\label{4.73}
 J_{11} = \gm_1 - a_{12} y \; , \qquad J_{12} = - a_{12}x \; , 
\qquad J_{21} = a_{21} y \; , \qquad J_{22} = - \gm_2 + a_{21} x \;  ,
\ee
which lead to the Jacobian trace and determinant 
$$
{\rm Tr}\hat J(x,y) = \gm_1 - \gm_2 + a_{21} x - a_{12} y \; ,
$$
$$
{\rm det}\hat J(x,y) =  (\gm_1 - a_{12} y) ( - \gm_2 + a_{21} y ) +
a_{12} a_{21} x y \; .
$$

At the trivial fixed point, we have
$$
{\rm Tr}\hat J(x_1^*,y_1^*) = \gm_1 - \gm_2 \; , \qquad
{\rm det}\hat J(x_1^*,y_1^*) =  - \gm_1 \gm_2 \; .
$$
The Jacobian eigenvalues are
\be
\label{4.74}
J_1(x_1^*,y_1^*) = \gm_1 \; , \qquad J_2(x_1^*,y_1^*) = - \gm_2 \;   .
\ee
Hence, the trivial fixed point is unstable, being a saddle point.

For the second fixed point, we get
$$
{\rm Tr}\hat J(x_2^*,y_2^*) = 0 \; , \qquad 
{\rm det}\hat J(x_2^*,y_2^*) = \gm_1 \gm_2 \;  ,
$$
hence the Jacobian eigenvalues become
\be
\label{4.75}
 J_1(x_2^*,y_2^*) = i\om \; , \qquad J_2(x_2^*,y_2^*) = - i\om \;   ,
\ee
where
\be
\label{4.76}
 \om = \sqrt{\gm_1 \gm_2 } \;  .
\ee

This means that the nontrivial fixed point is an elliptic point for any nonzero 
parameters of the evolution equation. The corresponding temporal behavior is given by 
permanent oscillations. The predators and prey oscillate with the same frequency $\om$, 
the predator curve being shifted to the right with respect to the prey curve.

There are several other types of equations describing population evolution. Some of 
which are mentioned below. 

\vskip 2mm

{\bf Singular Malthus Equation}

\vskip 2mm

The Malthus equation (\ref{4.53}) can be modified \cite{Foerster_52} to the form
\be
\label{4.77}
 \frac{dN}{dt} = \gm N^m \qquad ( m \geq 1) \;  ,   
\ee
whose solution diverges by power law
\be
\label{4.78}
 N = \frac{1}{ (\ep\gm)^{1/\ep}(t_c - t)^{1/\ep} } \;  ,
\ee
if $\varepsilon = m - 1 > 0$ and the critical time is
\be
\label{4.79}
 t_c = \frac{1}{N(0)^\ep \ep\gm} \;  .
\ee
For $m=1$, we return to the original Malthus equation that yields an exponentially 
divergent solution at $t\ra\infty$.

Such strongly singular solutions were applied to rationalize the super-exponential 
growth of the human world population, financial markets, material failures, 
earthquakes, climate changes, and dynamics of other systems (see \cite{Yukalov_53}).

\vskip 2mm

{\bf Generalized Lotka-Volterra Model} 

\vskip 2mm

The Lotka-Volterra model can also be generalized to the case of multiple species for 
the populations $N_i$, with $i = 1,2,\ldots$, yielding
\be
\label{4.80}
 \frac{dN_i}{dt} =\left( \gm_i + \sum_j A_{ij} N_j\right) N_i \;  .
\ee
The dimensionless populations can be introduced by normalizing $N_i$, for example,
to the total initial population $N(0) = \sum_i N_i(0)$, which gives the reduced form
\be
\label{4.81}
\frac{dx_i}{dt} =\left( \gm_i + \sum_j a_{ij} x_j\right) x_i \; ,
\ee
in which
$$
 x_i \equiv \frac{N_i}{N(0)} \; , \qquad a_{ij} \equiv A_{ij} N(0) \; .
$$
The signs of the parameters $\gamma_i$ and $a_{ij}$ can be different, providing 
a large variety of possible solutions \cite{Hofbauer_54}. For two species, we 
return to the original Lotka-Volterra model. 

\vskip 2mm

{\bf Predator-Prey Kolmogorov Model}

\vskip 2mm

A particular form generalizing the Lotka-Volterra Model, in which
\be
\label{4.82}
  \frac{dN_1}{dt} = f_1(N_1,N_2) \; N_1 \; , \qquad
  \frac{dN_2}{dt} = f_2(N_1,N_2) \; N_2 \; ,
\ee
under the conditions
\be
\label{4.83}
 \frac{\prt f_1}{\prt N_2} > 0 \; , \qquad
 \frac{\prt f_2}{\prt N_1} < 0 \; ,
\ee
is called the predator-prey Kolmogorov model \cite{Freedman_55,Brauer_56}. This model
can again be rewritten in the reduced form
\be
\label{4.84}
 \frac{dx_1}{dt} = f_1(x_1,x_2) \; x_1 \; , \qquad
  \frac{dx_2}{dt} = f_2(x_1,x_2) \; x_2 \; .
\ee
To model concrete cases, it is required to specify the functions $f_1$ and $f_2$.  

\vskip 2mm

{\bf Jacob-Monod Equations}

\vskip 2mm

The Jacob-Monod equations describe a single type of population $N_1$ that is being 
fed on the nutrient of amount $N_2$, 
\be
\label{4.85}
 \frac{dN_1}{dt} = f(N_2) \; N_1 \; , \qquad
  \frac{dN_2}{dt} = - \gm f(N_2) \; N_1 \; \;  .
\ee
For instance, this can be bacteria, playing the role of predators that are fed on 
a nutrient playing the role of the prey. The nutrient is getting depleted being 
consumed by the predators. At the same time, the nutrient is supplied into the 
system from outside, which is described by the function $f(N_2)$. The supply function 
is taken in the form, such that the nutrient becomes depleted ($N_2\ra 0$), as time 
increases, while the bacteria population reaches a fixed point characterizing a 
stationary value \cite{May_57}.

\vskip 2mm

{\bf Hutchinson Delayed Equation}

\vskip 2mm

There are as well several generalizations of the logistic equation (\ref{4.64}). 
Thus \cite{Hutchinson_58} considered an effective reproductive rate delayed in time, 
which gives the equation
\be
\label{4.86}
 \frac{dN(t)}{dt} = \gm \left[ \; 
1 \; - \; \frac{N(t-\tau)}{K} \; \right] \; N(t) \; .
\ee
Here $K$ is a fixed carrying capacity and $\tau$ is a delay time. The solution 
to this equation displays oscillations above the logistic curve.

\vskip 2mm

{\bf Multiple-Processes Delayed Equation}

\vskip 2mm

There exist many variants of the delay logistic equations 
\cite{Gopalsamy_1992,Kolmanovskii_60} designed for the description of single-species 
population $N$. For example, one can introduce multiple carrying capacities $K_j$ 
and multiple delay times $\tau_j$ characterizing different processes in the dynamics 
of a population. The single-species population dynamics, with multiple processes 
reads as
\be
\label{4.87}
  \frac{dN(t)}{dt} = \gm \left[ \; 
1  - \sum_j \frac{N(t-\tau_j)}{K_j} \; \right] \; N(t) \;  .
\ee

\vskip 2mm

{\bf Peschel-Mende Hyperlogistic Equation}

\vskip 2mm

The accelerated growth of population, such that exists for the human world 
population, can be described \cite{Peschel_61} by means of two additional 
powers $m$ and $n$, which leads to the equation
\be
\label{4.88}
  \frac{dN}{dt} = \gm \left( 1 \; - \;\frac{N}{K} \right)^n \; N^m \;  .
\ee
This equation leads to a solution that can be fitted well to the world population 
dynamics for some finite intervals of time. The solution to this equation is similar
to a modified sigmoid curve, which does not exceed the given carrying capacity $K$. 
The Verhulst logistic equation (\ref{4.64}) is recovered for $m = 1$ and $n = 1$.

\vskip 2mm

{\bf Hyperlogistic Delayed Equations}

\vskip 2mm

The Peschel-Mende hyperlogistic equation can be extended to the delayed equation 
\cite{Haberl_62}
\be
\label{4.89}
  \frac{dN(t)}{dt} = \gm \left[ \; 1 \; - 
\;\frac{N(t-\tau)}{K} \right]^n \; N^m(t) \;  .
\ee
This equation can also be treated as an extension of the Hutchinson delayed equation 
(\ref{4.86}). The solution to this delayed equation describes a population that can 
exceed the fixed carrying capacity $K$ for some finite period of time, although finally 
it decreases below $K$.

\subsection{Replicator Equation}

A very popular equation employed for modeling the evolution of different types of 
social and biological systems is the replicator equation describing population dynamics,
including dynamics in genetic theory and in evolution game theory. The replicator 
equation characterizes the ensemble of populations $N_i$ of different species, each 
being endowed with a fitness $f_i$. It is possible to consider either discrete or 
continuous in time processes. The continuous replicator equation is
\be
\label{4.90}
\frac{dN_i}{dt} = ( f_i - \overline f ) N_i \; ,
\ee
where 
\be
\label{4.91}
\overline f \equiv \frac{1}{N} \sum_i f_i N_i
\ee
is the average society fitness. The total population 
\be
\label{4.92}
N = \sum_i N_i = const
\ee
is assumed to be constant, hence the equation is defined on a simplex \cite{Hofbauer_54}. 

In terms of the reduced fractions
\be
\label{4.93}
 x_i \equiv \frac{N_i}{N} \qquad \left( \sum_i x_i = 1 \right) \;  ,
\ee
the equation takes the form
\be
\label{4.94}
 \frac{dx_i}{dt} = ( f_i - \overline f ) x_i \;  ,
\ee
with the average fitness
\be
\label{4.95}
 \overline f \equiv \sum_i f_i x_i \; .
\ee
This equation leads to a conclusion that in a society of several species only the
species with a largest fitness survives. 

To illustrate how the equation works, let us consider the simple case of two 
populations, $N = 2$, so that $x_1 + x_2 = 1$. Then the system of two equations 
can be reduced to a single equation, say
\be
\label{4.96}
 \frac{dx_1}{dt} = ( f_1 - f_2) ( 1 - x_1) x_1 \;  .
\ee
There are two fixed points, $x_1^* = 0$ and $x_1^* = 1$. The Jacobian is
\be
\label{4.97}
 J(x_1) = ( f_1 - f_2) ( 1 - 2x_1) \;  .
\ee
At the fixed points, this gives
\be
\label{4.98}
 J(0) = f_1 - f_2 \; , \qquad J(1) = - (f_1 - f_2 ) \;  .
\ee
Therefore, if the fitness of the first species $f_1$ is larger than that of the second 
species, $f_2$, then the stable fixed point corresponds to
\be
\label{4.99}
 x_1^* = 1 \qquad ( f_1 > f_2 ) \;  .
\ee 
However, when the fitnesses are such that $f_1$ is smaller than $f_2$, then only the
second species survives, since
\be
\label{4.100}
 x_1^* = 0 \qquad ( f_1 < f_2 ) \;   .
\ee
This result one often formulates as a sentence ``the strongest survives".

\subsection{Free Replicator Equation}

It is interesting that the same qualitative result follows from the replicator 
equation that is free from the normalization conditions (\ref{4.92}) or (\ref{4.93}),
which can be called free replicator equation. Then we need to deal with the number 
of equations equal to the number of species.

Thus, for the case of two species, we have two equations
\be
\label{4.101}
  \frac{dx_1}{dt} = ( f_1 - x_1 f_1 - x_2 f_2 ) x_1 \; , \qquad
  \frac{dx_2}{dt} = ( f_2 - x_1 f_1 - x_2 f_2 ) x_2 \; .
\ee
We assume that the species fitnesses are different, $f_1 \neq f_2$, otherwise the
species could not be distinguished. 

There are three fixed points, where either
\be
\label{4.102}
 x_1^* = 0 \; , \qquad  x_2^* = 0 \;  ,
\ee
or
\be
\label{4.103}
  x_1^* = 0 \; , \qquad  x_2^* = 1 \;  ,
\ee
or 
\be
\label{4.104}
  x_1^* = 1 \; , \qquad  x_2^* = 0 \;  .
\ee
The Jacobian matrix has the components
$$
J_{11}(x_1,x_2) = ( 1 - 2x_1) f_1 - x_2 f_2 \; , 
\qquad J_{12}(x_1,x_2)  = - x_1 f_2 \; ,
$$
\be
\label{4.105}
J_{21}(x_1,x_2)  = - x_2 f_1 \; , \qquad
 J_{22}(x_1,x_2) = ( 1 - 2x_2) f_2 - x_1 f_1 \;  .
\ee

At the fixed point (\ref{4.102}), we have
\be
\label{4.106}
J_{11}(0,0)  =  f_1 \; , \qquad J_{12}(0,0) = 0 \; , \qquad
 J_{21}(0,0) = 0 \; , \qquad J_{22}(0,0) = f_2  \; .
\ee
This gives the positive Lyapunov exponents
\be
\label{4.107}
\lbd_1 = f_1 \; , \qquad \lbd_2 = f_2 \; ,
\ee
showing that this fixed point is not stable. 

At the fixed point (\ref{4.103}), the Jacobian components are
\be
\label{4.108}   
J_{11}(0,1)  =  f_1 - f_2 \; , \qquad J_{12}(0,1) = 0 \; , \qquad
 J_{21}(0,1) = -f_1 \; , \qquad J_{22}(0,1) = - f_2  \;    ,
\ee
which gives the Lyapunov exponents
\be
\label{4.109} 
 \lbd_1 = f_1 - f_2  \; , \qquad \lbd_2 = - f_2 \;   .
\ee
Hence the fixed point (\ref{4.103}) is stable when $f_1 < f_2$. Then only the second
species survives.

At the fixed point (\ref{4.104}), for the Jacobian we get
\be
\label{4.110}
J_{11}(1,0)  =  - f_1  \; , \qquad J_{12}(1,0) = -f_2 \; , \qquad
 J_{21}(1,0) = 0 \; , \qquad J_{22}(1,0) =  f_2 - f_1 \;    ,
\ee 
thence the Laypunov exponents are
\be
\label{4.111}
 \lbd_1 = f_2 - f_1  \; , \qquad \lbd_2 = - f_1 \;    .
\ee
This tells us that the fixed point (\ref{4.104}) is stable provided that $f_1 > f_2$, 
when the second species dies out, while the first species survives.

\subsection{Influence of Noise}

The evolution equations considered above are called deterministic, since, given the
form of an equation and initial conditions, the following dynamics is uniquely defined.
In realistic situations, it may happen that the evolution of a society is subject to
random perturbations, termed random noise. Then, instead of Eq. (\ref{4.7}), one comes
to the equation
\be
\label{4.112}
\frac{dx}{dt} = f(x,t) + g(x,t) \xi(t) 
\ee
containing, in addition to $f(x,t)$, a term describing the action of an external noise.
Usually, setting
\be
\label{4.113}
 \xi(t)\; dt = d W(t) \; ,
\ee
one writes (\ref{4.112}) in the form
\be
\label{4.114}
 dx = f(x,t) \; dx + g(x,t) \; dW(t) \;  ,
\ee
since the noise term is often modeled by the Gaussian white noise that is not 
differentiable. In the case of white noise, the variable $W(t)$ characterizes a 
standard Wiener process. For time in the interval $[0,t]$, the random variable $W(t)$ 
is drawn from the normal law with zero mean and the standard deviation $\sqrt{t}$. 
The equation (\ref{4.114}) is called stochastic differential equation (see the books 
\cite{Kloeden_96,Gardiner_97,Lindenberg_98,Honerkamp_99}). The quantity $f(x,t)$ is 
called drift term and $g(x,t)$, diffusion term. For a while, we consider a single 
variable $x = x(t)$. The generalization to many variables will be given at the end 
of the section.  

In practical calculations, the dynamics of $x(t)$, in the presence of noise, is 
usually represented through the Euler-Maruyama scheme \cite{Maruyama_100}. For this 
purpose, the interval $[0,t]$ is partitioned in discrete points $t_n = n \Delta t$, 
with $n = 0,1,2,\ldots$. Then equation (\ref{4.114}) is integrated between $t_n$ and 
$t_{n+1} = t_n + \Delta t$, which gives
\be
\label{4.115}
 x(t_{n+1}) - x(t_n) = \int_{t_n}^{t_n+ \Dlt t} f(x,t) \; dt +
\int_{t_n}^{t_n+ \Dlt t} g(x,t) \; d W(t) \;  .
\ee
The first integral in the right-hand side is a usual Riemann integral that, for 
small $\Delta t$, can be written as
\be
\label{4.116}
 \int_{t_n}^{t_n+ \Dlt t} f(x,t) \; dt \cong f(x(t_n),t_n) \Dlt t \;   .
\ee

The Riemann integral does not depend on the choice of the variable $t$ inside the 
interval $[t_n, t_n + \Delta t]$, so that instead of $t_n$ in the right-hand side 
of the above formula it is possible to take any $t_n' \in [t_n, t_n + \Delta t]$. 
However, the integral over a stochastic variable depends on this choice. There exist 
two accepted ways of choosing the points of $t$ in the second integral of (\ref{4.115}), 
following Stratonovich or Ito \cite{Kloeden_96,Gardiner_97,Lindenberg_98,Honerkamp_99}). 
According to Stratonovich, for infinitesimally small $\Delta t$, one takes 
$[g(x(t_n + \Delta),t_n + \Delta) + g(x(t_n),t_n)]/2$, while according to Ito one has 
to take $g(x(t_n),t_n)$. When employing the Euler-Maruyama numerical scheme 
\cite{Maruyama_100} one uses the Ito representation. Then the second integral in 
(\ref{4.115}), understood in the sense of Ito, because of which it is called 
the Ito integral, writes as
\be
\label{4.117}
\int_{t_n}^{t_n+ \Dlt t} g(x,t) \; d W(t) = g(x(t_n),t_n) \; \left[ \; 
W(t_n+ \Dlt t) - W(t_n) \; \right ] \;  .
\ee
The random variable $W(t_n)$ is drawn from the normal distribution with zero mean and 
standard deviation $\sqrt{\Delta t}$. For each $t_1 < t_2$, the normal random variable
$W(t_2) - W(t_1)$ is independent of the random variable $W(t_1)$. The difference can 
be represented as
\be
\label{4.118}
W(t_n + \Dlt t) - W(t_n) = R(t_n) \; \sqrt{\Dlt t} \; ,
\ee
where $R(t_n)$ is drawn from the normal distribution with zero mean and standard 
deviation one. Thus we come to the iterative equation
\be
\label{4.119}
x(t_{n+1}) = x(t_n) + f(x(t_n),t_n) \Dlt t +
g(x(t_n),t_n) \; R(t_n) \; \sqrt{\Dlt t} \;   .
\ee
In those cases, where the diffusion term is constant, $g(x,t)=\sigma$, we have
\be
\label{4.120}
 x(t_{n+1}) = x(t_n) + f(x(t_n),t_n) \Dlt t + \sgm \; R(t_n) \; \sqrt{\Dlt t} \; .
\ee
 
Under the situation with several variables $x_i$, where $i=1,2,\ldots$, and several 
noise terms, one has
\be
\label{4.121}
  dx_i = f_i(x_i,t) \; dt + \sum_j g_{ij}(x,t) \; dW_j(t) \; .
\ee
For a diagonal diffusion matrix $g_{ij}(x,t)=\sgm_i\dlt_{ij}$, this reduces to
\be
\label{4.122}
x_i(t_{n+1}) = x_i(t_n) + f_i(x(t_n),t_n) \Dlt t + 
\sgm_i R(t_n) \; \sqrt{\Dlt t} \;   .
\ee

The existence of noise superimposes on a dynamical trajectory random deviations, whose 
amplitude depends on the value of $\sigma$.

\subsection{Fokker-Planck Equation}

In the presence of noise, the system trajectory is not uniquely defined, but there 
exists a set of possible trajectories depending on the particular realization of the
random noise. For the stochastic differential equation (\ref{4.114}), the probability 
density of $x$ is described by the Fokker-Planck equation \cite{Risken_101}
\be 
\label{4.123}
\frac{\prt}{\prt t} \; p(x,t) = - \;
\frac{\prt}{\prt x} \; \left[ \; f(x,t) p(x,t) \; \right] +
\frac{\prt^2}{\prt x^2} \left[ \; D(x,t) p(x,t) \; \right] \; ,
\ee
with the diffusion coefficient
\be
\label{4.124}
 D(x,t) \equiv \frac{1}{2} \; g^2(x,t) \;  .
\ee
Introducing the probability current
\be
\label{4.125}
j(x,t) \equiv f(x,t) p(x,t) \; - \; 
\frac{\prt}{\prt x} \; \left[ \; D(x,t) p(x,t) \; \right] 
\ee
yields the continuity equation
\be
\label{4.126}
 \frac{\prt}{\prt t} \; p(x,t) + \frac{\prt}{\prt x} \; j(x,t) = 0 \;  .
\ee

One often is interested in the stationary solution to the Fokker-Planck equation, 
provided it exists. A stationary solution can arise as time tends to infinity, 
so that the functions $f(x,t)$, $g(x,t)$, and $D(x,t)$ tend to time-independent
expressions $f(x)$, $g(x)$, and $D(x)$. Respectively, the probability density 
tends to $p(x)$, satisfying the equation 
\be
\label{4.127}
 \frac{d}{dx} \; j(x) = 0 \;  ,
\ee
in which the stationary probability current is
\be 
\label{4.128}
 j(x) = f(x) p(x) \; - \; \frac{d}{dx} \; [\;  D(x) p(x) \; ] \;  .
\ee
The latter equation implies that
\be
\label{4.129}
 j(x) = const  .
\ee

Suppose that the quantity $x$ varies in an interval starting at $x_0$, so that 
$x \geq x_0$.
Assuming the reflecting boundary condition
\be
\label{4.130}
 j(x_0) = 0 \;  , 
\ee
we get
\be
\label{4.131}
  j(x) = 0 \;  ,
\ee
which is equivalent to the equation
\be
\label{4.132}
  f(x) p(x) \; - \; \frac{d}{dx} \; [\;  D(x) p(x) \; ] = 0\;  .
\ee
The solution to the latter equation is
\be
\label{4.133}
 p(x) = \frac{C}{D(x)} \; \exp\left\{
\int_{x_0}^x \frac{f(x')}{D(x')} \; dx' \right\} \;  ,
\ee
with the constant $C$ defined from the normalization condition
\be
\label{4.134}
  \int p(x) \; dx = 1 \; ,
\ee
where the integration is over the whole range of variation of $x$. 

Expression (\ref{4.133}) is named potential solution, as far as it can be rewritten
in the form
\be
\label{4.135}
 p(x) = C \exp \{ - U(x) \} \;  ,
\ee
with the potential
\be
\label{4.136}
  U(x) = \ln D(x)  - \int_{x_0}^x \frac{f(x')}{D(x')} \; dx' \; .
\ee

All above formulas can be easily generalized to a multivariate case, where the 
Fokker-Planck equation reads as
\be
\label{4.137}
  \frac{\prt}{\prt t} \; p(x,t) = - \sum_i \frac{\prt}{\prt x_i} \;
[ \; f_i(x,t) p(x,t) \; ] +
\sum_{ij} \frac{\prt^2}{\prt x_i\prt x_j} \; [ \; D_{ij}(x,t) p(x,t) \; ] ,
\ee
with the diffusion matrix of the elements
\be
\label{4.138}
 D_{ij}(x,t) = \frac{1}{2} \sum_k g_{ik}(x,t) g_{jk}(x,t) \;  .
\ee

\section{Generalized Evolution Equations}

Usually, in the evolution equations the rates of change and population interactions 
are treated as constant parameters. In more realistic, and hence more general, 
situations they may be functions of the population variables. Below, we consider 
several such generalized equations.

\subsection{Functional Carrying Capacity}

In addition to direct interactions of different populations, there exist 
indirect interactions through the influence of the populations on the mutual 
carrying capacities. This concerns as well the influence of populations on their 
own carrying capacity. 

For example, the carrying capacity of a human society depends on the activity of 
humans. The technological evolution of humans has allowed increase of effective
carrying capacity for humans. At the same time, humans can destroy their carrying 
capacity, e.g. by destroying their habitat. 

To formulate the equations with a functional carrying capacity, let us start with 
the standard form of an evolution equation
\be
\label{5.1}
\frac{dN_i}{dt} = \left( \gm_i + \sum_j A_{ij} N_j\right) N_i + \Phi_i\;  ,
\ee
where $\gamma_i$ is an effective birth rate for a population, if $\gamma_i$ is 
positive, or a death rate, when $\gamma_i$ is negative. If one considers not 
population evolution, but dynamics of production of a firm or like that, then 
a positive $\gamma_i$ takes the sense of gain rate, while a negative $\gamma_i$ 
means loss rate. The quantity $A_{ij}$ describes effective interactions between 
the $i$-th and $j$-th populations. Usually, $A_{ij}$ is treated as a constant 
or, sometimes, as a given function fixed by external forces 
\cite{Begon_63,Ricklefs_64,Freedman_65}.

Generally, in the population interactions it is admissible to distinguish direct 
interactions and indirect interactions through the mutual influence on their 
carrying capacities \cite{Yukalov_53,Yukalov_66}. Therefore, the effective 
interactions can be represented in the form
\be
\label{5.2}
A_{ij} = - \; \frac{B_{ij}}{K_j}  \qquad ( K_j > 0 ) \;   ,
\ee
with $B_{ij}$ being a direct interaction parameter and $K_j$, carrying capacity 
of the $j$-th population. The influence of populations on the carrying capacity 
is assumed, which implies that it depends on these populations:
\be
\label{5.3}  
K_j = K_j(N_1(t-\tau_1), N_2(t-\tau_2), \ldots ) \; .
\ee
The populations entering $K_j$ are delayed, since to induce a change in the carrying 
capacity requires time. The term $\Phi_i$ is caused by the influx of populations from 
outside, if any.

As always, it is more convenient to deal with reduced dimensionless quantities, for 
which we introduce 
\be
\label{5.4}
 x_i \equiv \frac{N_i}{N_0} \; , \qquad \vp_i \equiv \frac{\Phi_i}{N_0} \;  .
\ee
Here the normalization quantity $N_0$, for a while, is not defined. It will be 
chosen later on so that to simplify the reduced equation. Then the evolution equation 
reads as
\be
\label{5.5}
 \frac{dx_i}{dt} = \left( 
\gm_i - \sum_j \frac{B_{ij}N_0}{K_j} \; x_j \right) x_i + 
\vp_i \;  .
\ee
If there are no external fluxes from outside, the term $\varphi_i$ should be omitted.

To illustrate the structure of the evolution equation, let us study the case with a 
single population without external flux. Then (\ref{5.5}) reduces to
\be
\label{5.6}
 \frac{dx}{dt} = \left( \gm \; - \; \frac{BN_0}{K} \; x \right) x \;  .
\ee
Using the sign function
\begin{eqnarray}
\nonumber
{\rm sgn} x \equiv \left\{ \begin{array}{rl}
-1 \; , ~ & ~ x < 0 \\
 0 \; , ~ & ~ x = 0 \\
 1 \; , ~ & ~ x > 0 \\
\end{array}
\right.
\end{eqnarray}
makes it possible to write
\be
\label{5.7}
 \gm = |\; \gm \; | {\rm sgn} \gm \; , \qquad 
 B = |\; B \; | {\rm sgn} B \;  .
\ee
Defining the dimensionless carrying capacity
\be
\label{5.8}
q \equiv \frac{K}{N_0} \; \left| \; \frac{\gm}{B} \; \right |
\ee
and measuring time in units of $\gamma^{-1}$ yields the evolution equation
\be
\label{5.9}
 \frac{dx}{dt} = ( {\rm sgn}\gm ) x - ( {\rm sgn} B ) \; \frac{x^2}{q} \;  .
\ee

Formally, this looks as a logistic equation. However the principal difference here 
is the functional dependence of the carrying capacity on the delayed population,
\be
\label{5.10}
  q = q(x(t-\tau) ) \; .
\ee
 
To find out the explicit form of the functional carrying capacity, let us assume 
that it can be expanded in powers of the population, so that
$$
K \simeq K_0 + K_1 N + K_2 N^2 + \ldots \; .
$$
This is equivalent to the expansion
\be
\label{5.11}
 q(x) \simeq q_0 + q_1 x + q_2 x^2 + \ldots \;  .
\ee
The carrying capacity has to retain its sense, hence to be finite for asymptotically 
small amount of population. This means that 
\be
\label{5.12}
q_0 = \frac{K_0}{N_0} \; \left| \; \frac{\gm}{B} \; \right | ~ > ~ 0 \;   .
\ee

To simplify the equation, we may choose the normalization quantity $N_0$ in the form
\be
\label{5.13}
 N_0 = K_0 \; \left| \; \frac{\gm}{B} \; \right | \;  ,
\ee
hence $q_0 = 1$, as a result of which the expansion of the carrying capacity becomes
\be
\label{5.14}
 q(x) \simeq 1 + b_1 x + b_2 x^2 + \ldots \;  .
\ee

When the influence of the population on the carrying capacity is small, it is 
possible to limit ourselves by the linear approximation $1 + b_1 x$, as has been 
considered in \cite{Yukalov_66}. However, this form, when used under sufficiently 
strong destructive influence of population on the carrying capacity, can lead to 
the appearance of zero in the denominator and the occurrence of unreasonable 
divergencies. 

To find a more general expression for the carrying capacity, valid under 
arbitrarily strong influence of populations, it is necessary to find an effective 
limit of expansion (\ref{5.11}). For this purpose, we can resort to the exponential 
summation guaranteeing a positive effective limit 
\cite{Yukalov_67,Yukalov_68,Gluzman_69}. This gives
\be
\label{5.15}
 q(x) = \exp\{ bx(t-\tau) \} \;  .
\ee

In this way, the evolution equation reduces to \cite{Yukalov_70}
\be
\label{5.16}
\frac{dx}{dt} = ({\rm sgn} \gm ) x - 
({\rm sgn } B ) x^2 \exp\{- bx(t - \tau) \} \;  .
\ee
The {\it production parameter} $b$ characterizes the influence of the population 
on the carrying capacity. The production parameter ($b>0$) is positive for the 
case of productive activity of population, creating additional means for survival.  
And the production parameter ($b < 0$) becomes negative when the population destroys 
the given carrying capacity. In such a situation, it is, actually, a destruction 
parameter. 

The initial condition for the delay equation is
$$
x(t) = x_0 \qquad ( t \leq 0 ) \;   .
$$

Depending on the signs of the parameters $\gamma$ and $B$, there can happen the 
following four different types of evolution models characterized by:
$$
{\rm sgn} \gm = 1 \; \qquad {\rm sgn} B = 1 \qquad ( gain + competition) \; ,
$$
$$
{\rm sgn} \gm = 1 \; \qquad {\rm sgn} B = -1 \qquad ( gain + cooperation) \; ,
$$
$$
{\rm sgn} \gm = - 1 \; \qquad {\rm sgn} B = 1 \qquad ( loss + competition) \; ,
$$
$$
{\rm sgn} \gm = -1 \; \qquad {\rm sgn} B = -1 \qquad ( loss + cooperation) \;   .
$$

\subsection{Evolutionary Stable States}

Delay equations allow for the Lyapunov stability analysis, similar to that for 
the standard differential equations \cite{Kolmanovskii_60}. The fixed points, or 
stationary states, are defined by the equation
\be
\label{5.17}
( {\rm sgn}\gm ) x^* - ({\rm sgn} B ) ( x^*)^2 \exp( - bx^*) = 0 \;   .
\ee
There always exists the trivial solution
\be
\label{5.18}
x_1^* = 0 \qquad ( - \infty < b < \infty ) 
\ee
existing for any ${\rm sgn} \gamma$ and ${\rm sgn} B$. 

Nontrivial solutions require the validity of the relation
\be
\label{5.19}
 \frac{ {\rm sgn}\gm}{ {\rm sgn} B } = x^* \exp(-bx^*) > 0 \;  ,   
\ee
which imposes the constraint
\be
\label{5.20}
  {\rm sgn}\gm =  {\rm sgn} B \qquad ( x^* > 0 ) \;  .
\ee
Under this condition, fixed points are given by the equation
\be
\label{5.21}
 x^* = \exp (bx^* ) \;  .
\ee

When the population destroys its carrying capacity, that is $b<0$, there can 
exist one fixed point in the range
\be
\label{5.22}
  0 < x_2^* \leq 1 \qquad ( b \leq 0 ) \; .
\ee
If the population increases its carrying capacity, hence $b>0$, but so that $b<1/e$,
there can occur two fixed points, one in the range
\be
\label{5.23}
  1 < x_2^* < e \qquad \left( 0 <  b < \frac{1}{e} \right) 
\ee
and the other for
\be
\label{5.24}
   x_3^* > e \qquad \left( 0 <  b < \frac{1}{e} \right)  .
\ee
At the bifurcation point $b = 1/e$, the fixed points $x_2^*$ and $x_3^*$ coincide:  
\be
\label{5.25}
 x_2^* = x_3^* = e \qquad \left( b = \frac{1}{e} \right) \;  .
\ee
There are no stationary nontrivial states for $b > 1/e$.

The fixed-point stability is characterized by the behavior of small deviations
from the fixed point. Substituting into the evolution equation (\ref{5.16}) the 
definition
\be
\label{5.26}
x = x^* + \dlt x
\ee
yields the linearized equation 
\be
\label{5.27}
\frac{d}{dt} \; \dlt x(t) = \left[ {\rm sgn} \gm - 2( {\rm sgn} B) x^* e^{-bx^*}
\right] \; \dlt x(t) + b ( {\rm sgn}\gm ) x^* \;\dlt x(t-\tau) \;   .
\ee

When the solution to (\ref{5.27}) is bounded, the solution to equation (\ref{5.16}) 
is Lyapunov stable. When the solution to (\ref{5.27}) converges to zero for 
$t \ra \infty$, a fixed point is asymptotically stable.

The trivial fixed point $x_1^* = 0$ is stable when
\be
\label{5.28}
{\rm sgn}\gm = - 1 \; , \qquad {\rm sgn} B = \pm 1 \; , \qquad
- \infty < b < \infty \; , \qquad \tau \geq 0 \; ,
\ee
but it is always unstable for ${\rm sgn}\gamma = 1$, 

The motion in the vicinity of nontrivial fixed points is described by equation
(\ref{5.27}), which, using relation (\ref{5.19}), reduces to the equation
\be
\label{5.29}
 \frac{d}{dt} \; \dlt x(t) = - ({\rm sgn}\gm ) \; \dlt x(t) +
b ( {\rm sgn}\gm ) x^* \; \dlt x(t-\tau) \;  .
\ee
Looking for the solution in the form
\be
\label{5.30}
  \dlt x(t) ~ \propto ~ e^{-\lbd t} \; ,
\ee
we get the equation for the Lyapunov exponent $\lambda$ depending on the studied
fixed point.

\subsection{Punctuated Evolution}

In the biological evolution theory there exists a hypothesis, called punctuated 
equilibrium, suggesting that the evolutional changes of biological species are marked 
by episodes of rapid speciation between long periods of little or no change. This
type of evolution, that occurs rapidly, being separated by periods of stasis, or 
equilibrium, is called {\it punctuated equilibrium} 
\cite{Gould_72,Eldredge_1985,Gould_2002}. If biological changes can be described by 
a quantitative characteristic, then the corresponding graph has the shape of a ladder.
A mathematical model of such a punctuated development can be represented by a delayed
equation \cite{Yukalov_70}. This model might characterize not only the evolution of
biological species, but also the evolution of firms and other organizations. In order 
to emphasize that this type of development is rather general, and can occur not only 
for biological species, it is called {\it punctuated evolution}.  

Let us consider the most realistic case of population characterized by gain and 
competition \cite{Yukalov_70}, where
\be
\label{5.31}
 {\rm sgn} \gm = {\rm sgn} B = 1 \;  .
\ee
Then equation (\ref{5.16}) reads as
\be
\label{5.32}
 \frac{dx}{dt} = x - x^2 \exp \{ - b x(t-\tau) \}\;  .
\ee

In that case, the trivial fixed point $x_1^* = 0$ is never stable. The nontrivial 
fixed point $x_2^*$ is stable when either
\be
\label{5.33}
0 < x_2^* \leq \frac{1}{e} \qquad ( b \leq -e \; , ~ \tau < \tau_2 ) \;   ,
\ee
where
\be
\label{5.34}
\tau_2 = \frac{1}{\sqrt{(bx_2^*)^2-1} } \; 
\arccos \left( \frac{1}{bx_2^*} \right) \; ,
\ee
or in the range
\be
\label{5.35}
 \frac{1}{e} < x_2^* < e \qquad 
\left( - e < b \leq \frac{1}{e} \; , ~ \tau > 0 \right) \;  .
\ee
The fixed point $x_3^* > e$ is always unstable under gain and competition. So that 
the sole stable fixed point is $x_2^* \in (0,e)$. 

When the production parameter $b$ is seminegative, which implies that the population
does not produce its carrying capacity but rather destroys it or, in the best case, 
retains the given capacity value, then the population remains bounded. More precisely, 
the following theorem takes place \cite{Yukalov_70}.

\vskip 2mm

{\bf Theorem on population boundedness}. {\it The solution $x(t)$ to the evolution
equation (\ref{5.32}), for $b \leq 0$, any finite $\tau \geq 0$, and any initial 
conditions $x_0 \geq 0$, is bounded for all times $t \geq 0$, and, for $b < 0$, 
there exists a time $t_0 = t_0(x_0,\tau)$ such that
$$
0 \leq x(t) \leq 1 \qquad ( t > t_0 ) \;  .
$$
The proof is given in} Ref. \cite{Yukalov_70}.

\vskip 2mm 
It is important to note that, for some values of the production parameter $b$, the 
basin of attraction of $x_2^*$ is not the whole positive semiline $x_0 \geq 0$, 
but a limited interval. This happens for $b \in (0, 1/e)$, when the basin of 
attraction is given by the inequalities
\be
\label{5.36}
0 < x_0 < x_3^* \qquad \left( 0 < b \leq \frac{1}{e} \right) \;   .
\ee
For the values of $b$ satisfying condition (\ref{5.36}), but with $x_0 > x_3^*$, 
the solution $x$ tends to infinity, as $t \ra \infty$.  

Overall, there exist the following regimes of population dynamics:

\vskip 2mm
(i) {\it Punctuated unbounded growth}. This growth happens when $b$ is outside 
of the stability region of $x_2^*$, so that
\be
\label{5.37}
 b > \frac{1}{e} \; , \qquad \tau > 0 \; , \qquad x_0 > 0 \;  .
\ee
An analogous unbounded punctuated growth happens when $b$ is inside the stability 
region, but $x_0$ is outside of the attraction basin of $x_2^*$, which occurs for
\be
\label{5.38}
 0 < b < \frac{1}{e} \; , \qquad \tau > 0 \; , \qquad x_0 > x_3^* \;   .
\ee

The unbounded punctuated behavior happens when either the production parameter is 
sufficiently large or the initial level of creative activity is enough high.

\vskip 2mm
(ii) {\it Punctuated convergence to a bounded state}. Punctuated convergence to
a finite stationary state $x_2^*$ happens for positive production parameters, when
\be
\label{5.39}
0 < b < \frac{1}{e} \; , \qquad \tau > 0 \; , \qquad x_0 < x_3^* \;   .
\ee
The solutions tend to the stationary state $x_2^*$ by punctuated steps from below, 
if $x_0 < x_2^*$, and from above, if $x_0 > x_2^*$.

\vskip 2mm
(iii) {\it Oscillatory convergence to a bounded state}. When the production parameter 
is negative, the approach to the stationary state becomes oscillatory. For
\be
\label{5.40}
 -e < b < 0 \; , \qquad \tau \geq  0 \; , \qquad x_0 > 0 \;  ,
\ee
there appear sharp reversals after almost horizontal plateaus. For
\be
\label{5.41}
 b < -e  \; , \qquad \tau \leq  \tau_2 \; , \qquad x_0 > 0 \;  ,
\ee
the population dynamics becomes strongly oscillating when approaching to the focus
$x_2^*$. 

\vskip 2mm
(iv) {\it Everlasting oscillations}. When the destructive action is rather strong
and the time delay is long, so that
\be
\label{5.42}
  b < -e  \; , \qquad \tau \geq  \tau_2 \; , \qquad x_0 > 0 \;   ,
\ee
then there develops the regime of everlasting oscillations. 

\vskip 2mm
The discussed regimes of dynamics under gain and competition are described in 
detail in Ref. \cite{Yukalov_70}, as well as other cases corresponding to gain 
and cooperation, loss and competition, and loss and cooperation. The punctuated 
evolution is typical only for the regime of gain and competition. This most 
interesting regime is rather widespread for many population societies. Thus it 
occurs in the evolution of biological species, growth of world population and 
urban population, development of science and technology, development of art and 
culture, energy production and consumption, growth of social organizations, growth 
of alive organisms, improvement of individual abilities, proliferation of cells 
and bacteria, as well as in the decay of biological organisms and societies 
\cite{Yukalov_53,Yukalov_66,Yukalov_70,Gould_72}.

\subsection{Symbiosis of Species}

The term symbiosis characterizes close and sufficiently long-time interactions between 
different biological species. In biological and ecological societies, symbiotic 
relationships are widespread. For example, the symbiosis between plant roots and 
fungi is a typical feature of numerous ecosystems. Co-existence of coral reefs and 
fishes is another well known example. Numerous other examples can be found in the 
books \cite{Boucher_73,Douglas_74,Sapp_75,Ahmadjian_76,Townsend_77}. In human 
societies, the examples of symbiosis are also widespread. Symbiotic relations 
are common for firms and banks, people and government, culture and language, economic 
and intellectual levels of society, basic science and technological applications.

\subsubsection{Interaction through Carrying Capacities}

The main idea in the new model of symbiosis is the observation that in symbiotic 
relations it is not the species themselves that interact directly with each other, 
but that symbiotic species influence the carrying capacities of each other. This 
implies that in the evolution equations of species the carrying capacities 
$K_j$ are functions of the populations $N_1$, $N_2$, etc. It is convenient to 
normalize the populations $N_i$ with respect to different normalization constants 
$M_i$, introducing the fractions
\be
\label{5.43}
 x_i \equiv \frac{N_i}{M_i} \qquad ( M_i = const ) \; .
\ee
Then the reduced form of the equations, without external forces, can be written 
as
\be
\label{5.44}
 \frac{dx_i}{dt} = \left( 
\gm_i - \sum_j \frac{B_{ij} M_j}{K_j} \; x_j \right) \; x_i \;  .
\ee
The convenience of using different normalizations is easily understood if one 
remembers that different species can exhibit rather different numbers of their 
members. For instance, the most important to humans symbiosis is that one between 
the human body and the numerous organisms of the microbiome, where there are 
more than $2000$ bacterial species and about $10^{14}$ microorganisms. The latter 
number is about ten times the number of cells of the human body \cite{Fasano_78}. 
Essentially different numbers of different populations can require different 
normalization constants. 

The coexisting species are characterized by direct interactions $B_{ij}$ and 
the interactions through their carrying capacities. In the general case, one 
can take into account the whole matrix of the interaction elements $B_{ij}$. 
When the direct interactions between the members of the same kind of species 
are much larger than between the members of different species, then the matrix 
$B_{ij}$ is approximately diagonal, 
\be
\label{5.45}
 B_{ij} \approx B_{ii} \dlt_{ij} \;  .
\ee
Therefore equation (\ref{5.44}) reduces to 
\be
\label{5.46}
 \frac{dx_i}{dt} = \gm_i \left( 
1 - \; \frac{B_{ii} M_i}{\gm_i K_i} \; x_i \right) \; x_i \;   .
\ee

It is possible to introduce dimensionless carrying capacities
\be
\label{5.47}
 q_i \equiv \frac{K_i}{K_i^0}     
\ee
that are functions $q_i=q_i(x_1,x_2,\ldots)$ such that $q(0,0,\ldots)=1$, 
which is possible by defining the appropriate normalization constants $K_i^0$. 
Generally, the variables entering $q_i$ should be delayed in time, such that 
$x_i=x_i(t-\tau_i)$.

Let us consider the case when the parameters $\gamma_i$ and $B_{ii}$ are positive. 
And let us choose the normalization factors $M_i$ as
\be
\label{5.48}
 M_i \equiv K_i^0 \; \frac{\gm_i}{B_{ii}} \;  .
\ee
Then we have
\be
\label{5.49}
 \frac{dx_i}{dt} = \gm_i \left( 1 -\; \frac{x_i}{q_i} 
\right) \; x_i \; .
\ee
 
Considering the case of two species, we define $x_1\equiv x$ and $x_2\equiv y$. 
Then equations (\ref{5.49}) reduce to the system of two equations
\be
\label{5.50}
 \frac{dx}{dt} = \gm_1 \left( 1 -\; \frac{x}{q_1} 
\right) \; x \; , \qquad
 \frac{dy}{dt} = \gm_2 \left( 1 -\; \frac{y}{q_2} 
\right) \; x \;.
\ee
Let us measure time in units of $1/\gamma_2$ and define the ratio
\be
\label{5.51}
 \al \equiv \frac{\gm_1}{\gm_2} \;  .
\ee
Thus we come to the equations
\be
\label{5.52}
\frac{dx}{dt} = \al \left( x - \; \frac{x^2}{q_1} \right) \; , 
\qquad 
\frac{dy}{dt} =  y - \; \frac{y^2}{q_2} \;  .
\ee

Now we need to define the carrying capacities $q_i$ that, generally, are the 
functions of the delayed populations
\be
\label{5.53}
 q_i = q_i(x_1(t-\tau_1),x_2(t-\tau_2), \ldots) \; .
\ee
For the case of two populations, $q_i=q_i(x,y)$. If we expand the function 
$q_i(x,y)$ in a series in powers of populations, we get the form as
\be
\label{5.54}
 q_i \simeq 1 + c_1 x + c_2 y + c_{11} x^2 + c_{22} y^2 +c_{12}xy +
\ldots \;   .
\ee

If the mutual influence of populations on the carrying capacities of each other 
is weak, then $q_i$ can be expanded over populations, with limiting ourselves 
by the lowest terms of the expansion \cite{Yukalov_53}. In the case of strong 
influence of symbiotic populations on each other, limiting ourselves by several 
first terms gives an expression that can become zero, thus producing spurious 
divergences in the terms containing $1/q_i$. To include in the consideration 
strong mutual influence, the expansions can be summed by means of exponential 
summation \cite{Yukalov_67,Yukalov_68,Gluzman_69}, thus avoiding spurious zeroes 
in the effective carrying capacity \cite{Yukalov_79,Yukalov_80}. In that way, we 
can derive the form
\be
\label{5.55}
 q_i = \exp\left\{ 
\sum_j b_{ij} x_j + \sum_{jk} c_{ijk} x_j x_k \right\} \;  .
\ee
The first term in the exponential describes the influence on the carrying 
capacity of separate populations not correlated with each other. The second 
term in the exponential characterizes the impact of the populations on the 
carrying capacity, when the symbiotic populations correlate with each other.  

From the general expression (\ref{5.55}), it is possible to set off two typical 
cases. One case is when the main terms in the exponential are those corresponding 
to the action of the symbiotic species without their mutual correlations, that is 
when the carrying capacities, for the case of two species, have the form
\be
\label{5.56}
 q_1 = e^{by} \; , \qquad q_2 = e^{gx} \qquad 
(uncorrelated \; symbiosis) \;  ,
\ee  
which can be called {\it uncorrelated symbiosis}, since the species, in the 
process of their action on the carrying capacities, do not correlate with each 
other. And the other situation describes {\it correlated symbiosis}, when the 
species influence the carrying capacities being mutually correlated, which for 
two species reads as
\be
\label{5.57}
 q_1 = e^{bxy} \; , \qquad q_2 = e^{gxy} \qquad 
(correlated \; symbiosis)  \;  .
\ee
The self-action of the species on their own carrying capacities is neglected 
here assuming that the influence of the symbiotic species is more important. 
It is also possible to study a mixed case when $q_1$ corresponds to a correlated 
symbiosis, while $q_2$, to uncorrelated symbiosis.  

Representing symbiosis through the mutual action of the symbiotic species on 
the carrying capacities of each other allows for the description of all types 
of symbiosis, which is straightforwardly connected with the signs of the 
symbiotic parameters $b$ and $g$. Below we give the classification of simbiotic 
types. In order to be precise and not to disturb the meaning, the formulation of 
the definitions below are given closely following Refs. \cite{Yukalov_79,Yukalov_80}. 

\vskip 2mm
(i) {\bf Mutualism} is a relationship between different species when both of 
them receive mutual benefit, which corresponds to the symbiotic parameters  
\be
\label{5.58}
 b > 0 \; , \qquad g > 0 \;  .
\ee

\vskip 2mm
(ii) {\bf Commensalism} is a relationship, when one of the species benefits from 
the coexistence with the other species, while the other one is neutral, getting 
neither profit nor harm, which is defined by one of the conditions
\be
\label{5.59}
b > 0 \; , \qquad g = 0 \;
\ee
\vskip 2mm
\be
\label{5.60}
b = 0 \; , \qquad g > 0 \;   .
\ee

\vskip 2mm
(iii) {\bf Parasitism} is a relation, when at least one of the coexisting species 
is harmful to the other one, which is characterized by one of the conditions
\be
\label{5.61}
b > 0 \; , \qquad g < 0 \;   ,
\ee
\vskip 2mm
\be
\label{5.62}
b = 0 \; , \qquad g < 0 \;   ,
\ee
\vskip 2mm
\be
\label{5.63}
 b < 0 \; , \qquad g < 0 \;  ,
\ee  
\vskip 2mm
\be
\label{5.64}
b < 0 \; , \qquad g = 0 \;  ,
\ee
\vskip 2mm
\be
\label{5.65}
b < 0 \; , \qquad g > 0 \;   .
\ee

As is seen, there exist various types of symbiosis described by the systems 
of differential equations that can be solved numerically \cite{Rheinboldt_81}. 
Since symbiosis is extremely widespread in nature, there are numerous particular 
examples of coexisting species among biological and ecological societies, 
including bacteria and viruses 
\cite{Boucher_73,Douglas_74,Sapp_75,Ahmadjian_76,Townsend_77,Ivanitsky_82}.

\subsubsection{Uncorrelated Symbiosis}

Dynamics of symbiotic species, under the uncorrelated symbiosis and the assumption 
of approximately equal rates $\gm_1$ and $\gm_2$, are described by the system of 
equations
\be
\label{5.66}
 \frac{dx}{dt} = x- x^2 e^{-by} \; , \qquad
 \frac{dy}{dt} = y- y^2 e^{-gx} \;  ,
\ee 
with the symbiotic parameters $b\in(-\infty,\infty)$ and $g\in(-\infty,\infty)$. 
For simplicity, we assume that the delay times of the populations entering the 
carrying capacities are very small, so that can be neglected. The initial 
conditions are $x_0 = x(0)$ and $y_0 = y(0)$. 

For any values of the parameters, there always exist three trivial fixed points, 
$\{x^*=0,y^*=0\}$, $\{x^*=1,y^*=0\}$, and $\{x^*=0, y^*= 1\}$, which are unstable 
for all $g$ and $b$. Nontrivial fixed points are given by the equations
\be
\label{5.67}
  x^* = e^{by^*} \; , \qquad y^* = e^{gx^*} \; , 
\ee
that can be rewritten as
\be
\label{5.68}
 x^* = \exp \left( be^{gx^*} \right) \; , \qquad 
 y^* = \exp \left( ge^{by^*} \right) \; .
\ee
The Lyapunov exponents are defined by the expressions
\be
\label{5.69}
\lbd_1 = - 1 + \sqrt{bgx^*y^*} \; , \qquad 
\lbd_2 = - 1 - \sqrt{bgx^*y^*} \;   .
\ee
  
Under {\it mutualism}, the analysis shows that there can exist the following 
regimes depending on the parameters $b>0$ and $g>0$:

\begin{enumerate}[label=(\roman*)]
\item   
Unbounded growth of populations with time. 

\item
Convergence to a stationary state.
\end{enumerate}

In the case of {\it parasitism}, the situation depends on whether a single species 
is parasitic or both species are parasites. When a single species is parasitic, 
that is when either $b<0$, while $g>0$ or $b>0$, while $g<0$, then only one regime 
exists, when the populations tend to their stationary states.  

When both species are parasitic, so that $b<0$ and $g<0$ then, depending on the 
symbiotic parameters, there can exist two regimes:

\begin{enumerate}[label=(\roman*)]
\item
Convergence to single stationary state.

\item
Bistability with two stationary states.
\end{enumerate}

The details can be found in Ref. \cite{Yukalov_80}.

\subsubsection{Correlated Symbiosis}

Correlated symbiosis of two species is characterized by the system of equations
\be
\label{5.70}
 \frac{dx}{dt} = x- x^2 e^{-bxy} \; , \qquad
 \frac{dy}{dt} = y- y^2 e^{-gxy} \;   ,
\ee
where we again assume that $\gamma_1$ is close to $\gamma_2$. Similarly to the 
previous case of uncorrelated symbiosis, there always exist three trivial fixed 
points, $\{x^*=0,y^*=0\}$, $\{x^*=1,y^* =0\}$, and $\{x^*=0,y^*=1\}$, which are 
unstable for any $g$ and $b$. Nontrivial fixed points are the solutions to the 
equations
\be
\label{5.71}
  x^* = e^{b x^*y^*} \; , \qquad y^* = e^{g x^*y^*} \;  ,
\ee
that can be rewritten as
\be
\label{5.72}
 x^* = \exp \left\{ b (x^*)^{1+g/b} \right\} \; , \qquad 
 y^* = \exp \left\{ g (y^*)^{1+b/g} \right\} \;  .
\ee
The Laypunov exponents are defined by the relations
\be
\label{5.73}
 \lbd_1 = - 1 \; , \qquad \lbd_2 = - 1  + (b+g) x^* y^* \; .
\ee

In the case of mutualism, where $b>0$ and $g>0$, there are the following types 
of behavior:

\begin{enumerate}[label=(\roman*)]
\item
Unbounded growth of both populations.

\item
Convergence of populations to stationary states.
\end{enumerate}

In the case of parasitism, when either one of the symbiotic parameters 
is negative, while the other is positive, or both parameters are negative, the 
following situations can happen:

\begin{enumerate}[label=(\roman*)]
\item
Convergence to stationary states.

\item
Unbounded growth of parasitic population and dying out of host population. 
\end{enumerate}

The detailed investigation is given in Ref. \cite{Yukalov_80}.

\subsubsection{Mixed Symbiosis}

In the previous examples describing symbiosis, it is possible to notice the 
symmetry corresponding to the interchange between the populations $x$ and $y$. 
It also may happen that two symbiotic species exhibit nonsymmetric relations 
that can be characterized by the case of mixed symbiosis, when one of the 
species displays correlated symbiosis, while the other species, uncorrelated 
symbiosis, This case is described by the equations
\be
\label{5.74}
 \frac{dx}{dt} = x- x^2 e^{-bxy} \; , \qquad
 \frac{dy}{dt} = y- y^2 e^{-gx} \;   .
\ee

Similarly to the previous cases, there always exist three trivial fixed points, 
$\{x^*= 0, y^*= 0\}$, $\{x^*= 1, y^*= 0\}$, and $\{x^*=0, y^*=1\}$, which are 
unstable for all symbiotic parameters $b$ and $g$. Nontrivial fixed points are 
given by the equations 
\be
\label{5.75}
 x^* = e^{b x^*y^*} \; , \qquad y^* = e^{g x^*} \;   
\ee
that can be represented as
\be
\label{5.76}
 x^* =  \exp\left\{ b x^* e^{gx^*} \right\} \; , 
\qquad
 y^* =  \exp\left\{ g (y^*)^{by^*/g} \right\} \;  .
\ee
The Lyapunov exponents are
$$
\lbd_1 = \frac{1}{2} \; \left[\; b x^* y^* - 2 +
x^* \; \sqrt{by^* ( 4g + by^*) } \; \right] \; , 
$$
\be
\label{5.77}
 \lbd_2 = \frac{1}{2} \; \left[\; b x^* y^* - 2 -
x^* \; \sqrt{by^* ( 4g + by^*) } \; \right] \;  .
\ee

The dynamics of the symbiotic populations strongly depends on whether the 
influence of the species $y$ on the carrying capacity of species $x$ is 
mutualistic or parasitic, which is described by the sign of the symbiotic 
parameter $b$. If the latter is negative $(b<0)$, there exists the sole regime, 
when both populations tend to their stationary limits. If $b>0$ and $g>0$, two 
regimes can happen:

\begin{enumerate}[label=(\roman*)]
\item
Unbounded growth of both populations.

\item
Convergence to stationary states. 
\end{enumerate}

When $b>0$, but the first species is parasitic, with $g<0$, then there can exist 
two situations:

\begin{enumerate}[label=(\roman*)]
\item
Convergence to stationary states.

\item
Everlasting oscillations.
\end{enumerate}

Details can be found in Ref. \cite{Yukalov_80}. Applications to financial markets, 
considering symbiosis between asset prices and bonds, resulting in periodically 
growing and collapsing bubbles, are analyzed in Ref. \cite{Yukalov_83}.

\subsection{Role of Growth Rates}

When the growth rates $\gamma_1$ and $\gamma_2$ are essentially different, it is 
necessary to consider the system of equations
\be
\label{5.78}
 \frac{dx}{dt} = \al \left( x - \; \frac{x^2}{q_1} \right) \; ,
\qquad 
 \frac{dy}{dt} =  y - \; \frac{y^2}{q_2} \;  ,
\ee
in which $\alpha \equiv \gamma_1/\gamma_2$ and time is measured in units of 
$\gamma_2^{-1}$. Without the loss of generality, it is possible to take 
$\alpha > 1$. In the case of uncorrelated symbiosis, we have
\be
\label{5.79}
 \frac{dx}{dt} = \al \left( x - x^2 e^{-by} \right) \; ,
\qquad 
 \frac{dy}{dt} =  y - y^2 e^{-gx} \;   .
\ee

There are three trivial fixed points: the unstable node $\{0, 0\}$, with the 
Lyapunov exponents $\lbd_1 = 1$ and $\lbd_2 = \al$; a saddle $\{1, 0\}$, with 
the Lyapunov exponents $\lbd_1=1$ and $\lbd_2=-\al$; and the saddle $\{0, 1\}$, 
with the Lyapunov exponents $\lbd_1 = -1$ and $\lbd_2 = \al$. The nontrivial 
fixed points do not depend on the value of $\alpha$ and are defined as above.
The related Lyapunov exponents are
$$
\lbd_1 = -\; \frac{1}{2} \; ( 1 + \al) + \; 
\sqrt{(1-\al)^2 + 4\al b g x^* y^* } \; , 
$$
\be
\label{5.80}
\lbd_2 = -\; \frac{1}{2} \; ( 1 + \al) - \; 
\sqrt{(1-\al)^2 + 4\al b g x^* y^* } \; .
\ee

Although the fixed points do not depend on the rate $\alpha$, but the Lyapunov 
exponents do depend on it, as well as the related basins of attraction also depend 
on $\al$, which hence influences the stability of fixed points \cite{Cross_84}. 
Therefore by varying $\alpha$ it is admissible to obtain different regimes of 
motion, hence to realize dynamic phase transitions by the sole variation of the 
growth rate, with keeping other parameters unchanged. 

A usual dynamic transition implies a qualitative change of dynamical behavior,
when system parameters reach a bifurcation point. Then the type of fixed points 
changes. However, there can happen a nonstandard dynamic transition, when a 
qualitative change of dynamical behavior occurs due to the variation of the growth 
rate. In that case, the fixed points do not change, remaining the same, while a 
sharp change in dynamical behavior happens because the growth rate shifts the 
boundary of the basins of attraction, so that the initial point of a trajectory, 
which was inside the attraction basin, moves outside of it \cite{Yukalov_85}.

\subsection{Self-Organized Society}

The evolution of biological societies, including human societies, is a principally 
important old problem studied in voluminous literature beginning with Darwin 
\cite{Darwin_86,Darwin_87}. Societies usually are structured into groups representing 
particular features or traits. As an example of group representatives, it is possible 
to mention collaborators and defectors. The collaborators cooperate with each other 
for the benefits of the whole society, while defectors, on the contrary, exploit 
it \cite{Perc_30,Perc_31,Jusup_32}.

The evolution of groups is usually studied on the basis of the replicator equation 
discussed in the above sections. As has been explained, if the society consists of two 
groups only, cooperators and defectors, the latter always outperform the former, 
so that the sole evolutionary stable state is the state where there are no cooperators, 
but there exist solely defectors. It is evident that for a closed self-organized 
society, where there are no unlimited resources supplied from somewhere outside, 
this conclusion is absurd, since defectors produce nothing and, being left alone, 
cannot survive.

Our aim is to consider a self-organized society, whose means of survival are produced 
inside the society itself. A closed self-organized society cannot exist being composed 
solely of defectors because they will have no means for survival. Sometimes, to correct 
this unrealistic conclusion, one introduces punishers. However, these also require means 
for their existence and cannot survive if nothing is produced.

\subsubsection{Trait Groups}

In the present section, a self-organized structured society is described composed of 
the trait groups representing four types of typical agents: cooperators, defectors, 
regulators, and outsiders. The principal novelty of the approach is that, instead 
of a single fitness or utility for each group, we introduce relative utilities for 
each group with respect to the society as a whole, and mutual utilities with respect 
to each other \cite{Yukalov_91}.

Let us consider a society composed of several groups, whose fractions are defined 
as the ratios
\be
\label{5.81}
x_i \equiv \frac{N_i}{N_0}
\ee
with respect to a normalizing number $N_0$ that can be chosen to define the initial 
total number of the society members
\be
\label{5.82}
N_0 = \sum_i N_i(0) \qquad ( t = 0 ) \;   .
\ee
Then, at the initial time there exists the normalization
\be
\label{5.83}
\sum_i x_i(0) = 1 \;  .
\ee
However, at the later times such a normalization, generally, is not valid. In general, 
the number of the society members can vary with time because of births, deaths, and 
the influx of outsiders. 

Biological, including human, societies have much in common with the structures of the 
organisms because of which the society groups can be straightforwardly compared 
with the parts of biological organisms. The classification below follows 
Refs. \cite{Yukalov_92,Sornette_93}.  

\begin{itemize}
\item
{\bf Cooperators} $(x_1)$, who contribute to the whole society. In a human society, 
the cooperators form the working force producing the gross domestic product. In a 
biological organism, cooperators can be associated with healthy cells.

\item
{\bf Defectors} $(x_2)$, who do not contribute to the society and can exist only 
owing to the work of cooperators. In a social system, the groups that benefit from 
the society support without contributing are prisoners, pensioners, and unemployed 
people. In a biological organism, defectors can be represented by ill cells. 

\item 
{\bf Regulators} $(x_3)$, who maintain order in the society and punish defectors 
and harmful outsiders. In a human society, this role is played by the police, the 
army, and the order enforcing bureaucracy. To support the existence of regulators, 
the society has to pay the necessary costs. In a biological organism, regulators 
can correspond to the cells of the immune system.

\item
{\bf Outsiders} $(x_4)$, who also exploit the society, but, contrary to defectors, 
the difference is that they enter the society from outside. The harmful outsiders 
could be interpreted as terrorists or as foreign invading armies. For biological 
organisms, outsiders could be pathogens or viruses infecting the organism.
\end{itemize}

As the evolution equations for fractions (\ref{5.81}), it is possible to accept the 
rate equations
\be
\label{5.84}
 \frac{dx_i}{dt} = 
\left( \gm_i + \sum_j a_{ij} x_j \right) \; x_i + \vp_i \;  .
\ee
These equations are formally equivalent to the rate equations (\ref{4.52}) or to
the Lotka-Volterra equations (\ref{4.81}). However, the meaning of the parameters 
entering equations (\ref{5.84}) is different. In the Lotka-Volterra equations,
$\gamma_i$ are the birth-death rates and $a_{ij}$ are intensities of interactions. 

In our case, each $\gamma_i$ plays the role of utility rate, or the rate of production
(or consumption), or the production of resources with respect to the whole society, and 
the parameters $a_{ij}$ are interpreted as the utility production (or consumption) by 
the $j$-th group with respect to the $i$-th group. Diagonal elements $a_{ii}$ correspond
to the competition between the members of the same group (hence $a_{ii} < 0$). The signs 
of non-diagonal $a_{ij}$ are defined depending on the usefulness of the group $j$ to 
group $i$, so that if $j$ is useful to $i$, then $a_{ij}$ is positive, while, when $j$
is harmful for $i$, then $a_{ij}$ is negative. External influx is assumed to exist 
only for outsiders. The definitions below are based on Refs. \cite{Yukalov_92,Sornette_93}.

\vskip 2mm
{\bf Cooperators}. They are useful for the society. Defectors are not useful for 
cooperators. Regulators require the support of cooperators which is a cost to the latter.
Harmful insiders also are not useful for cooperators. Summarizing this, we have:
\be
\label{5.85}
 \gm_1 > 0 \; , \qquad \vp_1 = 0 \; , \qquad a_{11} < 0 \; , \qquad
a_{12} < 0 \; , \qquad a_{13} < 0 \; , \qquad a_{14} < 0 \; .
\ee
 
\vskip 2mm  
{\bf Defectors}. They are not useful for the society. Cooperators are necessary for 
defectors who live at the expense of the former. Regulators suppress and punish 
defectors. Invaders are not useful for defectors. Thus:
\be
\label{5.86}
\gm_2 < 0 \; , \qquad \vp_2 = 0 \; , \qquad a_{21} > 0 \; , \qquad
a_{22} < 0 \; , \qquad a_{23} < 0 \; , \qquad a_{24} < 0 \;  .
\ee

{\bf Regulators}. They do not produce resources. Society needs to support regulators. 
Cooperators are necessary for regulators. The role of regulators is to maintain order 
and to punish defectors, whose presence justifies the existence of regulators. 
Similarly, regulators suppress harmful invaders, which justifies the existence of 
regulators. Therefore:
\be
\label{5.87}
\gm_3 < 0 \; , \qquad \vp_3 = 0 \; , \qquad a_{31} > 0\; , \qquad
a_{32} > 0 \; , \qquad a_{33} < 0 \; , \qquad a_{34} > 0 \; .
\ee

\vskip 2mm
{\bf Outsiders}. They are not useful for the society. But cooperators are necessary 
for outsiders. Outsiders exploit defectors by taking a part of their share, and 
benefit from their presence. Regulators, suppressing outsiders, are not useful to them. 
Hence:
\be
\label{5.88}
 \gm_4 < 0 \; , \qquad \vp_4 \geq 0 \; , \qquad a_{41} > 0 \; , \qquad
a_{42} > 0\; , \qquad a_{43} < 0 \; , \qquad a_{44} < 0 \; .
\ee

Taking into account the signs of the parameters, equations (\ref{5.84}) write as
$$
\frac{dx_1}{dt} = ( \gm_1 - |\; a_{11} \; | x_1 - |\; a_{12} \; | x_2
- |\; a_{13} \; | x_3 - |\; a_{14} \; | x_4 ) \; x_1 \; ,
$$
$$
\frac{dx_2}{dt} = ( - |\gm_2| + |\; a_{21} \; | x_1 - |\; a_{22} \; | x_2
- |\; a_{23} \; | x_3 - |\; a_{24} \; | x_4 ) \; x_2 \; ,
$$
$$
\frac{dx_3}{dt} = ( - |\gm_3| + |\; a_{31} \; | x_1 + |\; a_{32} \; | x_2
- |\; a_{33} \; | x_3 + |\; a_{34} \; | x_4 ) \; x_3 \; ,
$$
\be
\label{5.89}
\frac{dx_4}{dt} = ( - |\gm_4| + |\; a_{41} \; | x_1 + |\; a_{42} \; | x_2
- |\; a_{43} \; | x_3 - |\; a_{44} \; | x_4 ) \; x_4 + \vp_4  \;  .
\ee

In general, the quantities $a_{ij}$ could be functions of populations, as 
in the considered above cases of retarded carrying capacity, resulting in 
punctuated evolution, and symbiotic relations through functional carrying 
capacities. However, we shall not complicate the situation and will treat 
$a_{ij}$ as parameters. To reduce the number of parameters, it is possible 
to employ relations existing between them, keeping in mind that, by accepted 
interpretation, $a_{ij}$ is the utility rate for a group $i$ of a group $j$.
For brief, below we shall name $a_{ij}$ simply as utilities.  

The cooperators are the sole group producing resources for the whole society. 
These resources are denoted through $\gm_1$. The cooperators compete for these 
resources, which is described by the value $|a_{11}|$. This utility is actually 
all that is produced by the cooperators, which implies the equality
\be
\label{5.90}
 |\; a_{11} \; | = \gm_1 \;  .
\ee
The diagonal elements $|a_{ii}|$ describe the strength of competition among the 
members of a group $i$ for the available for them resources $\gamma_i$ playing 
the role of a carrying capacity. The standard dependence for the term 
characterizing competition on the carrying capacity is the inverse dependence
\be
\label{5.91}
 |\; a_{ii} \; | = \frac{C}{|\gm_i|} \; ,
\ee
where $C$ is a constant. From relation (\ref{5.90}) it follows that $C=\gm_1^2$.
Therefore
\be
\label{5.92}
 |\; a_{ii} \; | = \frac{\gm_1^2}{|\gm_i|} \;  .
\ee
The non-diagonal elements $|a_{ij}|$ play the role of mutual utility for the 
groups $i$ and $j$. The widely used expression for the mutual utility is the 
Bernoulli-Nash mutual utility \cite{Acocella_94,Moulin_95} that can be written 
as 
\be
\label{5.93}
 |\; a_{ij} \; | = \sqrt{|\gm_i \gm_j|} \qquad ( i \neq j) \; .
\ee
Also, there should exist the reciprocity relation
\be
\label{5.94}
a_{ij} = - a_{ji}   \qquad ( i \neq j)
\ee
signifying that, if a group $i$ receives some utility from a group $j$, then the 
group $j$ looses the same amount of utility.

We introduce the dimensionless parameters, describing the fractions of the wealth 
consumed by the related groups with respect to the total amount of resources 
produced by the society cooperators
\be
\label{5.95}   
a \equiv \frac{|\gm_2|}{\gm_1} \; , \qquad 
b \equiv \frac{|\gm_3|}{\gm_1} \; , \qquad
c \equiv \frac{|\gm_4|}{\gm_1} \; , 
\ee
and the dimensionless influx of outsiders
\be
\label{5.96}
\vp \equiv \frac{\vp_4}{\gm_1} \;   .
\ee

We measure time in units of $\gamma_1^{-1}$. Then equations (\ref{5.89}) acquire 
the form of the system of equations
\be
\label{5.97}
\frac{dx_i}{dt} = f_i \qquad ( i = 1,2,3,4 ) \;  ,
\ee
with the right-hand sides
$$
f_1 = 
\left( 1 - x_1 - \sqrt{a}\; x_2 -  \sqrt{b}\; x_3 -  \sqrt{c}\; x_4 
\right)  x_1  \; ,
$$
$$
f_2 = 
\left( -a + \sqrt{a}\; x_1 - \frac{1}{2} \; x_2 - \sqrt{ab}\; x_3 - 
\sqrt{ac}\; x_4 \right)  x_2  \; ,
$$
$$
f_3 = 
\left( -b + \sqrt{b}\; x_1 + \sqrt{ab} \; x_2 - \frac{1}{b}\; x_3 + 
\sqrt{bc}\; x_4 \right)  x_3  \; ,
$$
\be
\label{5.98}
f_4 = 
\left( -c + \sqrt{c}\; x_1 + \sqrt{ac} \; x_2 -  \sqrt{bc}\; x_3 - 
\frac{1}{c}\; x_4 \right)  x_4  \; .
\ee
   
One more reasonable restriction that should be accepted is the {\it conservation 
law} telling that one cannot consume more than it is produced. That is, what is 
consumed by defectors, regulators, and outsiders, cannot be larger than what is 
produced by cooperators,
\be
\label{5.99}
  |\gm_2 + \gm_3 + \gm_4| \leq \gm_1 \;  .
\ee
In dimensionless units this is equivalent to the inequality
\be
\label{5.100}
 a + b + c \leq 1 \;  .
\ee
If this condition becomes invalid, this means that defectors, regulators, and 
outsiders consume more than cooperators produce. This could be realized only under 
a supply from outside or by the destruction of the given carrying capacity, which 
implies an unstable situation.

\subsubsection{Coexistence of Cooperators and Defectors}

It is instructive to consider the coexistence of two typical groups, cooperators
and defectors. Recall that in the case of replicator equation, the stable state 
corresponds to the absence of cooperators and all society being composed of defectors.
As has been discussed, this absolutely unrealistic conclusion stems from the fact that
the replicator equation does not describe a self-organized society. Let us see what is
the situation in our case of a self-organized society. 

Considering cooperators and defectors, we have the evolution equations
\be
\label{5.101}
 \frac{dx_1}{dt} = \left( 1 - x_1 - \sqrt{a} \; x_2 \right)  x_1 \; ,
\qquad  
\frac{dx_2}{dt} = \left( - a + \sqrt{a} \; x_1 - \frac{1}{a}  x_2 
\right) x_2 \; .
\ee
Looking for fixed points satisfying conditions (\ref{5.100}), we find the sole 
solution
\be
\label{5.102}
  x_1^* = \frac{1+a^{5/2}}{1+a^2} \; , \qquad
  x_2^* = \frac{1 - \sqrt{a} }{1+a^2} \; a^{3/2} \; ,
\ee
that is stable under the condition
\be
\label{5.103}
 0 \leq a \leq 1 \;  .
\ee
From here, we find that the minimal fraction of cooperators occurs when
\be
\label{5.104}
\min_a x_1^* = 0.940 \qquad ( a = 0.565) \;   ,
\ee
while the maximal fraction of defectors is
\be
\label{5.105}
\max_a x_2^* = 0.083 \qquad ( a = 0.480 ) \;   .
\ee
Thus in a stable self-organized society, the amount of defectors cannot be 
larger than around $10\%$. If this amount is essentially larger, the society 
is not stable.

\subsection{Three Coexisting Groups}

The standard situation is the coexistence of three groups, cooperators, defectors, 
and regulators, while outsiders are not numerous, hence can be neglected. Then the
equations are
$$
\frac{dx_1}{dt} = \left( 1 - x_1 - \sqrt{a} \; x_2 - \sqrt{b} \; x_3 
\right)  x_1 \; ,
$$
$$
\frac{dx_2}{dt} = \left( - a + \sqrt{a} \; x_1 - \frac{1}{a} \; x_2 -
\sqrt{ab} \; x_3 \right)  x_2 \; ,
$$
\be
\label{5.106}
\frac{dx_3}{dt} = \left( - b + \sqrt{b} \; x_1 +
\sqrt{ab} \; x_2 - \; \frac{1}{b} \; x_3 \right) x_3 \; .
\ee
There is the sole fixed point satisfying the conservation law (\ref{5.100}),
$$
x_1^* = \frac{1+a^2b^2 + a^{5/2}(1+b^2)+b^{5/2}(1-a^2)}{(1+a^2)(1+b^2)} \; ,
$$
\be
\label{5.107}
x_2^* = 
\frac{1 - \sqrt{a}\; (1+b^2) - b^2 (1-2\sqrt{b})}{(1+a^2)(1+b^2)}\; a^{3/2} \; ,
\qquad
x_3^* = \frac{1-\sqrt{b}}{1+b^2}\; b^{3/2} \; ,
\ee
which is stable when
\be
\label{5.108}
0 \leq a < \left( \frac{2b^2\sqrt{b} - b^2 + 1}{1+b^2} \right)^2 \; ,
\qquad 0 \leq b \leq 1 \;  .
\ee
This shows that the minimal fraction of cooperators is
\be
\label{5.109}
\min_{a,b} x_1^* \approx 0.91 \qquad ( a \approx 0.45, ~ b \approx 0.55) \; ,
\ee
the maximal fraction of defectors is
\be
\label{5.110}
\max_{a,b} x_2^* \approx 0.08 \qquad ( a \approx 0.48, ~ b = 0) \;   ,
\ee
and the maximal fraction of regulators is
\be
\label{5.111}
\max_{a,b} x_3^* \approx 0.08 \qquad ( 0 \leq a \leq 0.52, ~ b = 0.48) \; .
\ee
Again we see that in a stable society the fractions of defectors and regulators
should not exceed about $10\%$ each.   

More discussions and applications of the theory for describing ant and bee colonies, 
and concrete countries is given in Ref. \cite{Yukalov_91}, where the influence of 
noise is also studied.

\section{Models of Financial Markets}

Financial markets can be treated as complex social systems, because of which their
behavior could allow for mathematical description. It is not our aim to plunge 
deeply into the ocean of economic theories, but our aim is to present some basic 
ideas the models of markets are based on and to illustrate them by simple examples.

\subsection{Efficient Market Model}

The old standing hypothesis, the functioning of financial markets is based on, is 
the {efficient market hypothesis}. It is generally accepted 
\cite{Courtault_2000,Franck_2012} that the idea of efficient market has been 
anticipated by Bachelier \cite{Bachelier_102} who compared the motion of market prices 
with the Brownian motion. According to Bachelier, ``past, present and even discounted 
future events are reflected in market price, but often show no apparent relation to 
price changes". Fama \cite{Fama_1965,Fama_1970} distinguished three forms of efficient 
market hypothesis, weak form, semi-strong form, and strong form. Weak form assumes 
that the stock prices indicate the present public market information, the past 
performance does not play any role, although the prices may not reflect new 
information that is not yet publicly available. Semi-strong form states that the 
stock prices reflect both the market and non-market public information and that 
prices adjust quickly to any new public information that becomes available. Strong 
form insists that prices reflect the entirety of both public and private information,
historical and new, as well as insider information. Summarizing, the strong form
of the efficient market hypothesis assumes that:

\begin{itemize}

\item
All prices on traded assets already reflect all past available information.

\item
All prices instantly adjust to any newly appearing information.

\item
All prices instantly reflect even hidden information.

\end{itemize}

An efficient market is assumed to be absolutely equilibrium, random, and stable.
It unpredictably fluctuates, so that nobody can consistently achieve returns in
excess of average market returns. The motion of stock prices is compared with
random walk typical of Brownian motion. 

Typical agents acting in a market are supposed to be rational. Any one particular 
agent can be wrong and make mistakes, but the market as a whole is always right.
Agents errors are random, so that they are averaged out. A typical, or representative
agent is the so-called {\it Homo Economicus} who:

\begin{itemize}

\item
Possesses all existing information necessary for the correct price evaluation.

\item
Can immediately process all existing information. 

\item
Makes objective unbiased conclusions based on the maximization of expected utility.

\end{itemize}

Let $p=p(t)$ represent asset price, stock price, or market index. Dynamics of $p$,
for an efficient market, is assumed to follow the geometric Brownian motion that 
satisfies the stochastic differential equation
\be
\label{6.1}
 d\; p(t) = [\; y \; dt + \sgm \; dW(t) \;] \; p(t) \;  ,
\ee
in which $y$ is a drift rate of the price, $\sigma$, diffusion coefficient, or 
market volatility of returns, and $W$ is a random Wiener process. Note that the 
above equation can be interpreted as a rate equation
\be
\label{6.2}
 \frac{dp(t)}{dt} = R(t) p(t) \;  ,
\ee
with a randomly fluctuating rate
\be
\label{6.3}
 R(t) = y + \sgm \xi(t) \;  ,
\ee
where $\xi(t)=dW(t)/dt$ is white noise. This model describes the market motion as
on average exponential, with superimposed random fluctuations. 

Without the noise, the solution would be
\be
\label{6.4}
 p(t) = p(0) e^{yt} \; .
\ee
While in the presence of noise, the numerical solution is given by the iterative
scheme
\be
\label{6.5}
  p(t_{n+1}) = 
[\; 1 + y \Dlt t + \sgm R(t_n) \sqrt{\Dlt t} \; ] \; p(t_n) \;  .
\ee

\subsection{Diffusion Price Model}

In the diffusion price model, the drift term $y$ in Eq. (\ref{6.1}) is treated not 
as a constant, but as a random variable, so that the market is described by the 
system of equations
\be
\label{6.6}
dp = ( y dt + \sgm_1 dW_1) p \; , \qquad 
dy = -\gm y dt + \sgm_2 dW_2 \;   ,
\ee
where $\gm>0$ is called market friction. The initial conditions are $p(0)$ and 
$y(0)$. When there is no noise, the equations are
\be
\label{6.7}
 \frac{dp}{dt} = yp \; , \qquad 
\frac{dy}{dt} = -\gm y \;  .
\ee
Then the price is given by the expression
\be
\label{6.8}
 p(t) = p(0) e^{y(0)/\gm} \exp\left\{ - \; 
\frac{y(0)}{\gm} \; e^{-\gm t} \right\} \;  .
\ee
With time, the drift term $y$ tends to zero, 
\be
\label{6.9}
y(t) = y(0) e^{-\gm t} ~ \rightarrow ~ 0 \qquad ( t \ra \infty) \;  ,
\ee
and the price tends to the value
\be
\label{6.10}
 p(t) \simeq p(0) e^{y(0)/\gm}  \qquad ( t \ra \infty)  
\ee
named the {\it fundamental price}. Thus the market price, it seems, should tend 
with time to equilibrium. Random noise induces only small fluctuations around the 
equilibrium fundamental price value.  

Recall that this conclusion is based on the assumption of market efficiency. 
However, the questions remain: Are markets efficient in the sense discussed above? 
Are they in an equilibrium state or at least tend to equilibrium? Are they 
sufficiently stable? Are typical agents, acting in markets, really rational?

\subsection{Herding Market Model}

It has been noticed long time ago that agents are not completely rational, but their 
rationality is bounded \cite{Simon_103}. The agents are always subject to hesitations, 
biases, superstitions, and other feelings \cite{Kahneman_104}. The agents cannot be 
absolutely rational, since:

\begin{itemize}
\item
They have numerous cognitive biases.

\item
They do not possess all necessary information.

\item
They are not able to accomplish instantaneous calculations.

\item
Actually, they do not maximize some expected utility or other functionals.

\end{itemize}

Generally speaking, all agents, to more or less extent, are subject to various 
emotions that can essentially influence their decisions. It is hardly possible to 
numerically evaluate the role of emotions for each particular agent, since decision 
making is in principle a random process \cite{Yukalov_105}, but the quantification 
of emotions during this process is admissible on aggregate level \cite{Yukalov_106}.  
 
In particular, the members of a society are strongly subject to the so-called 
herding effect. As Poincar\'{e} \cite{Poincare_107} has written, ``Individuals 
who are close to each other, as they are in a market, do not take independent 
decisions $-$ they watch each other and herd." The herd instinct acts similarly 
to feedback field in nonlinear systems, which in a financial market can lead to 
strong fluctuations \cite{Friedman_108,Plummer_109,Haugen_110} and even can 
trigger business cycles \cite{Gordon_111,Schumpeter_112,Hall_113}. In addition, 
a market is not compulsorily equilibrium due to government interference, 
different external manipulations, corruption, insider trading etc. Financial 
crises, accompanied by strong fluctuations, as booms and crashes, can be caused 
by herding effect \cite{Sornette_2003}.  

The behavior of markets cannot be always stationary. Actually, sometimes markets 
look as almost stationary and efficient, but sometimes they become strongly 
nonequilibrium and inefficient. The market boom-crash fluctuations remind us the 
so-called heterophase fluctuations in statistical systems, where these fluctuations 
are caused by collective effects \cite{Yukalov_15,Yukalov_36}. To make the market 
behavior richer, so that to include in its description regime switching between 
conventions and business cycles, it is necessary to take account of collective 
effects \cite{Yukalov_115}. 

Let us define the logarithmic {\it mispricing}
\be
\label{6.11}
 x \equiv \log \; \frac{p}{p^*} \;  ,
\ee
where the price $p$ is normalized to a fundamental price $p^*$ and the logarithm 
can be taken with respect to any convenient base. The equation for the mispricing 
\be
\label{6.12}
dx = y dt + \sgm_1 dW
\ee
follows from the diffusion price model, as in the previous section. However the 
drift term is assumed to satisfy the stochastic equation
\be
\label{6.13}
dy = f dt + \sgm_2 dW   ,
\ee
with its own drift force $f = f(x,y)$. As usual, $W$ is a Wiener process.  

The drift force $f(x,y)$ can be modeled by expanding $f(x,y)$ in powers of the 
variables $x$ and $y$ and then resorting to exponential summation \cite{Yukalov_68}, 
taking into account that the force $f(x,y)$ has to be antisymmetric with respect 
to the replacement $x \mapsto -x$ and $y \mapsto -y$. The resulting drift force
takes the form
\be
\label{6.14}
 f(x,y) = \al x + \bt y + Ax^3 \exp\left( - \; \frac{x^2}{\mu^2} \right) 
+ B y^3 \exp \left( - \; \frac{y^2}{\lbd^2} \right) \;  .
\ee
The meaning of the terms in force (\ref{6.14}) is as follows.

The term $\al x$ describes the individual response to changing price, with $\al<0$ 
being correcting response, while $\alpha > 0$ being speculative response. The term 
$\beta y$ characterizes the individual response to the changing price drift, with 
$\beta < 0$ being correcting response named market friction. 

The term
$$
Ax^3 \exp\left( - \; \frac{x^2}{\mu^2} \right)
$$
describes collective response to changing price, with $A < 0$ corresponding to 
correcting response and $A > 0$, to speculative response. The parameter $\mu$ 
defines the measure of uncertainty in the price value. The absence of uncertainty 
implies a fully informed market, when $\mu \ra 0$, hence 
\be
\label{6.15}
Ax^3 \exp\left( - \; \frac{x^2}{\mu^2} \right) ~ \longrightarrow ~ 0
\qquad ( \mu \ra 0 ) \;  .
\ee
The herding behavior corresponds to $A > 0$.

Collective response to the varying price drift is modeled by the term
$$
 B y^3 \exp \left( - \; \frac{y^2}{\lbd^2} \right) \; .
$$
Here $B < 0$ corresponds to contrarian response, while $B > 0$, to speculative 
response. The parameter $\lambda$ measures the level of market freedom, or 
deregulation. If a market is over-regulated, with $\lambda \ra 0$, then there is
no collective response,
\be
\label{6.16}
 B y^3 \exp \left( - \; \frac{y^2}{\lbd^2} \right)  ~ \longrightarrow ~ 0
\qquad ( \lbd \ra 0 ) \;  .
\ee
Herding occurs under $B > 0$. 

The phase portrait, defining $y = y(x)$, is given by the equation 
\be
\label{6.17}
 \frac{dy}{dt} = \frac{f(x,y)}{y} \;  .
\ee

Depending on the relative value of the model parameters, there exists a rich variety 
of different regimes describing markets with coexisting equilibrium, conventions, 
and business cycles. The inclusion of noise defines stochastic dynamics that is 
characterized by nonlinear geometric random walk processes with spontaneous regime 
shifts between different conventions and business cycles. This model provides a 
natural framework to explain dynamic regime shifts between different market states. 
These shifts are the result of the interplay between the individual and herding 
effects, under the presence of noise \cite{Yukalov_115}. The change of dynamic 
regimes induced by external noise leads to noise-induced dynamic transitions 
\cite{Horsthemke_116}.

\subsection{Time Series Analysis}

Financial and economic time series are a particular kind of time series, whose 
analysis is necessary in many applications. The standard methods of their analysis
can be found in the books \cite{Box_117,Bails_118,Granger_119}. Here we present a
method that has been developed rather recently \cite{Yukalov_120}. This method is
based on self-similar approximation theory (see reviews \cite{Yukalov_121,Yukalov_122})
interpreting the transfer between approximants in approximation space as the motion
of a dynamical system, with the approximant order playing the role of discrete time.  
The idea of the approach is the assumption that in given data, filtered from  noise, 
there exists a hidden law of evolution, which gives the possibility of understanding 
the future system development with a finite forecast horizon.

Suppose we observe a series of financial or economic market data expressed in the 
form of numbers, e.g. values or market indices, resulting in a set of data $z_i$ 
at the related moments of time $t_i$. It is necessary to keep in mind that the raw 
data as such are not representative for characterizing the market dynamics, since 
these data may contain a great deal of noise. Therefore there is no clear convincing 
reason to analyze time series data in precisely the form in which they are provided 
\cite{Granger_119}. The existence of noise is typical for practically all large 
systems, whether in physics or economics. Even equilibrium systems can exhibit rather 
strong fluctuations (see, e.g. \cite{Yukalov_123}).   

In markets, generally, there can happen two kinds of noise, caused by exogenous 
or endogenous sources. The former can be due to asset and wealth shocks caused 
by wars or other disasters, abrupt population changes, and like that \cite{Phelps_124}. 
Endogenous noise is produced by the system itself. The evolution of an economic system 
is essentially a disturbance of equilibrium in the economy. Within each economic 
system there always exists a source of energy sufficient for disrupting any 
equilibrium \cite{Andersen_125}. A long-term equilibrium cannot be reached, but 
fluctuations and noise are everlasting \cite{Hayek_126,Rothbard_127}. There exist 
different ways of filtering out noise \cite{Box_117,Bails_118,Granger_119}. The 
simplest and rather reasonable way is as follows. 

\vskip 2mm
{\bf Step 1}. Let us separate the studied time period into several time intervals, 
say $k + 1$ intervals, enumerated by the index $n = 0,1,2, \ldots, k$, with the 
corresponding market data sets,
\be
\label{6.18}
 \{ z_i^{(n)} , t_i^{(n)} : ~ i = 1,2,\ldots M_n \}  \qquad
( n = 0 , 1,2, \ldots, k) \;  .
\ee
Depending on the situation under study, each interval can be a day or a week or 
what is more appropriate. 

The values $z_i^{(n)}$ are subject to random noise. This noise can be smoothed 
out by composing, for each time interval $n$, the average values 
\be
\label{6.19}
x_n \equiv \frac{1}{M_n} \sum_{i=1}^{M_n} z_i^{(n)} \; . 
\ee
The value $x_n$ can be ascribed to a point of time $t_n$ inside the $n$-th time 
interval, which, e.g., can be chosen as the average of $t_i^{(n)}$, or as the left 
or right time point of the time interval in (\ref{6.18}), or, if the time intervals 
are equal to $\Delta t$, as $t_n=n\Dlt t$. Then we obtain the smoothed data set of 
the averages
\be
\label{6.20}
\mathbb{D}_k = \{ x_n , t_n : ~ n = 0,1,2,\ldots k \} \; .
\ee

\vskip 2mm
{\bf Step 2}. Assume that the data set (\ref{6.20}) represents a function $f(t)$ 
passing through the given time points. The most complete representation of $f(t)$ 
is given by the polynomial \cite{Lorentz_128,Pindyck_129}
\be
\label{6.21}
 f_k(t) = \sum_{n=0}^k a_n t^n = 
a_0 \left( 1 + \sum_{n=1}^k \frac{a_n}{a_0} \; t^n \right) \;  ,
\ee
such that 
\be
\label{6.22}
 f_k(t_n) = x_n \qquad ( n = 0,1,2,\ldots, k) \;  ,
\ee
which defines $a_n = a_n(x_1,x_2,\ldots,x_k)$.

\vskip 2mm
{\bf Step 3}. The polynomial (\ref{6.21}) can be extrapolated beyond the time $t_k$
by employing self-similar approximation theory \cite{Yukalov_121,Yukalov_122}, using 
self-similar exponential approximants \cite{Yukalov_67,Yukalov_68}. Then we obtain
the approximants of first order
\be
\label{6.23}
 f_1^*(t) = a_0 \exp ( c_1 t) \;  ,
\ee
where $c_1$ is defined by comparing the expansion in powers of $t$ of (\ref{6.23}) 
with the first order series (\ref{6.21}), which gives
\be
\label{6.24}
 c_1 = \frac{a_1}{a_0} \;  .
\ee

The second-order approximant reads as
\be
\label{6.25}
 f_2^*(t) = a_0 \exp ( c_1 t \exp ( c_2 t \tau_2 ) ) \;  ,
\ee
with the control function $\tau_2$ defined by the fixed-point equation
\be
\label{6.26}
 \tau_2 = \exp(c_2 t \tau_2 ) \;  ,
\ee
and with $c_2$ following from the accuracy-through-order procedure by equating 
the expansion of approximant (\ref{6.25}), under $\tau_2 = 1$, with polynomial 
(\ref{6.21}), which leads to   
\be
\label{6.27}
c_2 = \frac{a_2}{a_1} \; - \; \frac{a_1}{2a_0} \; .
\ee

The third-order approximant is
\be
\label{6.28}
 f_3^*(t) = a_0 \exp ( c_1 t \exp ( c_2 t \exp(c_3 t \tau_3 ) ) ) \;  ,
\ee
with the parameter $c_3$ defined by the accuracy-trough order procedure,
\be
\label{6.29}
c_3 = \frac{a_3}{a_1c_2} \; - \; \frac{c_2}{2} \; - c - \; 
\frac{c_1^2}{6 c_2} \;   , 
\ee
and the control function $\tau_3$ prescribed by the fixed-point condition 
\be
\label{6.30}
  \tau_3 = \exp(c_3 t \tau_3 ) \;  .
\ee

Continuing this procedure, we get 
\be
\label{6.31}
f_k^*(t) = 
a_0 \exp ( c_1 t \exp ( c_2 t \tau_2 \ldots \exp(c_k t \tau_k ) ) \ldots ) \; ,
\ee
with the control functions
\be
\label{6.32}
 \tau_k = \exp(c_k t \tau_k ) \;    .
\ee 
Generally, control functions $\tau_k = \tau_k(t)$ can be defined by fixed-point 
conditions \cite{Yukalov_130}, as above, or by minimizing a cost functional 
\cite{Yukalov_131}.    

\vskip 2mm
Extrapolating the approximants (\ref{6.31}) to time $t > t_k$ gives nonlinear 
forecasts that can describe many booms and crashes which cannot be explained by 
standard methods. Detailed description of a number of examples can be found in 
Refs. \cite{Yukalov_130,Gluzman_132,Gluzman_133,Gluzman_134}. 

It is necessary to keep in mind that market bubbles and crashes can be provoked 
by many causes. There are indefinitely large number of functional relationships 
that can lead to such unstable effects as bubbles and crashes 
\cite{Sornette_2003,Rosser_135}. The standard time-series autoregressive methods 
are based on presumed structural stability and linearity, because of which they 
cannot in principle predict unstable nonlinear, and strongly nonequilibrium market 
movements, such as booms and crashes \cite{Zhang_136}. Markets are largely 
influenced by human emotions resulting in herding behavior
\cite{Zhang_136,Turner_137,Woods_138}. In the words of Woods \cite{Woods_138}, 
``It is easier to forecast weather than to predict stock market prices", and 
the essence of the standard autoregressive models is ``garbage in, garbage out".  

The exponential extrapolation, described above, is highly nonlinear and gives 
the hope for the possibility of grasping nonequilibrium market dynamics. However, 
since the origin and nature of different booms and crashes can be rather different, 
it may happen that it is necessary to employ different variants of self-similar 
extrapolation, for instance, different conditions defining control functions.

\section{Conclusion}

The content of this article is based on the lectures that have been given by the 
author during several years at the Swiss Federal Institute of Technology in Z\"{u}rich
(ETH Z\"{u}rich). The first part \cite{Yukalov_0} considered equilibrium social systems. 
In this part of the lectures, nonequilibrium social systems are discussed. In addition 
to the general information, several novel evolution equations are studied describing 
such phenomena as punctuated evolution, symbiosis of species with interactions through 
functional carrying capacity, dynamic phase transitions caused by the variation of 
growth rates, stability of self-organized societies, and herding in markets.  

The whole theme of social physics is huge and, of course, it is impossible to cover 
it in a single survey. The reader can find a lot more of models in the cited literature. 
Here the choice of the touched topics is motivated by the interests of the author.

\vskip 2mm

{\bf Funding}: This research received no external funding.

\vskip 2mm

{\bf Acknowledgments}: I appreciate very much numerous discussions with D. Sornette,
with whom several results mentioned above have been obtained. I am grateful for 
discussions and help to E.P. Yukalova.    
  
\vskip 2mm

{\bf Conflicts of Interest}: The author declares no conflict of interest.

\end{document}